\documentclass[aps, prb, amssymb, amsmath, 
twocolumn] {revtex4-2}

\usepackage{chngcntr}

\usepackage{graphicx}
\usepackage{subfigure}
\usepackage{color}
\usepackage[dvipsnames]{xcolor}
\usepackage{amsmath}
\usepackage{enumitem}
\usepackage{amssymb}
\usepackage[hidelinks]{hyperref}
\usepackage{cancel}
\usepackage{ulem}
\usepackage{multirow}

\newcommand{\be}{\begin{equation}}
	\newcommand{\ee}{\end{equation}}

\newcommand{\bea}{\begin{eqnarray}}
	\newcommand{\eea}{\end{eqnarray}}

\makeatother

\begin{document}

\preprint{APS/123-QED}

\setlength{\abovecaptionskip}{-60pt}

\title{
Disorder-induced spin-cluster magnetism in a doped kagome spin liquid candidate}

\author{
Arnab Seth$^1$, Joseph C. Prestigiacomo$^2$,  Aini Xu$^3$, Zhenyuan Zeng$^{3,4}$, Trevor D. Ford$^5$, B.S. Shivaram$^{5}$, Shiliang Li$^{3,4,6}$, Patrick A. Lee$^{7}$, and Itamar Kimchi$^1$ }

\affiliation{} 

\affiliation{$^1$School of Physics, Georgia Institute of Technology, Atlanta, Georgia 30332, USA.}

\affiliation{$^2$Materials Science and Technology Division, Naval Research Laboratory, Washington, DC 20375, USA.}

\affiliation{$^3$Beijing National Laboratory for Condensed Matter Physics,
Institute of Physics, Chinese Academy of Sciences, Beijing 100190, China.}

\affiliation{$^4$School of Physical Sciences, University of Chinese Academy of Sciences, Beijing 100190, China.}

\affiliation{$^5$Department of Physics, University of Virginia, Charlottesville, Virginia 22904, USA.}

\affiliation{$^6$Songshan Lake Materials Laboratory, Dongguan, Guangdong 523808, China}

\affiliation{$^7$Department of Physics, Massachusetts Institute of Technology, Cambridge, Massachusetts 02139, USA.}

\date{January 20, 2026}

\begin{abstract}
The search for new quantum spin liquid materials relies on systems with strong frustration such as spins on an ideal kagome lattice. However, lattice imperfections can have substantial effects which are as yet not well understood. In recent work, the two-dimensional  kagome system YCu$_3$(OH)$_6$[(Cl$_x$Br$_{(1-x)}$)$_{3-y}$(OH)$_y$]  has emerged as a leading candidate hosting a Dirac spin liquid which appears to survive at least for $x<0.4$, associated with alternating-bond hexagon (ABH) disorder. Here in magnetic samples with $x=0.58, y=0.1$ we report unusual in-plane ferromagnetic canting (FM) of the in-plane antiferromagnet (AFM), with an unusually wide regime of short-ranged order, and propose theoretical models to explain this behavior. First, we show that Kitaev type exchanges naturally arise on the kagome lattice to second order in the known Dzyaloshinskii-Moriya exchanges, and that these interactions can produce the unusual in-plane FM canting from antichiral AFM. Second, we propose a phenomenological model of weakly-FM-canted spin clusters to describe the short-ranged regime and analyze quantum fluctuations in an ABH toy model to show how ABH disorder can stabilize this regime. 
  The combination of experimental observation and theory suggests that kagome-Kitaev interactions and ABH disorder are necessary for describing the magnetic fluctuations in this family of materials, with potential implications for the proposed proximate spin liquid phase. 
\begin{description}
\item[PACS numbers]
75.30.Mb, 71.27.+a, 75.25.Dk
\end{description}
\end{abstract}

\maketitle

\section{Introduction}

Recently a new class of  quantum kagome antiferromagnet (AFM)  materials has been explored that minimize accidental substitutional disorder: YCu$_3$(OH)$_6$Cl$_3$ \cite{NormanRMP2016, ZorkoPRB2019, sunJCHEM2016}. Site mixing of magnetic and non-magnetic atoms is suppressed by their large difference in ionic radii. Rather than the sought after quantum spin liquid state this compound  appears to harbor a magnetic phase transition \cite{ZorkoPRB2019negative, ArhPRL2020} into the antichiral ordered AFM with 120$^\circ$ spin structure and negative vector chirality.  Since this system is well described by the two-dimensional (2D) nearest-neighbor kagome AFM model (with Heisenberg exchange of order 100 K) where no magnetic transition is expected, the magnetic order is likely created by additional out-of-plane Dzyaloshinskii-Moriya (DM) interaction \cite{ BernuPRB2020,Messio2010,Cepas2008,HeringPRB2017}. A partial substitution of Cl by O, giving the compound YCu$_3$(OH)$_6$O$_x$Cl$_{(3-x)}$ with x=1/3, results in the absence of static magnetic order \cite{BarthelemyPRM2019}. However, it is not clear what is the role of the O substitution, and in particular what is the correct microscopic description of the system and the substitution-induced disorder.  

More recently a related family of compounds has been discovered, and  some members of the family are shown not to order down to 50 mK. This family involves two doping parameters. First, the  Cl in the kagome compound can be replaced by a combination of Br and OH  \cite{ChenJMMM2020, HongPRB2022, Lunature2022,  ZengPRB2022, suetsugu_gapless_2024},  resulting in YCu$_3$(OH)$_6$Br$_2$[Br$_{1-y}$(OH)$_y$],
which we will refer to as YCOB. 
Second, YCOB is expanded into a family of compounds via partial replacement of Br by Cl, leading to YCu$_3$(OH)$_6$[(Cl$_x$Br$_{(1-x)}$)$_{3-y}$(OH)$_y$], which we denote as YCOB-Cl~\cite{Xu2023,shivaramPRB2024}.
Extensive studies of the heat capacity $C$ and linear magnetic susceptibility $\chi$ in high quality single crystals of this disordered system for low $x$ show behavior such as $C/T$ with linear-$T$ behavior at low $T$, that increases in magnetic fields $H$ with a linear-$H$ term beyond a threshold field, \cite{ZengPRB2022, Liu2022, Zengpreprintneutron} consistent with a Dirac spin liquid \cite{RanPRL2007, HermelePRB2008}.

These compounds are reported to host a special type of disorder, where OH substitutes the Br sites randomly above and below the kagome hexagons formed by Cu atoms \cite{Liu2022, Xu2023}. This leads to alternating pattern of spin exchange interaction strength around that hexagon, resulting in {alternating-bond hexagon} (ABH) defects. 
Similar kind of disorder is found in the Y-kapellasite Y$_3$Cu$_9$(OH)$_{19}$Cl$_8$ \cite{Puphal2017,BarthelemyPRM2019,ChatterjeePRB2023}. It is not clear how ABH disorder impacts the magnetic properties of YCOB-Cl.

In this joint theory-experiment work we report on two unusual magnetic features seen experimentally in  magnetic samples with $x=0.58, y=0.1$, and proceed to describe these features using a combination of theoretical models. The two puzzling features observed in experiment (Sec.~II) are as follows. First, single-crystal magnetization measurements show magnetic anisotropy that points to weak in-plane ferromagnetism (FM) beyond the dominant in-plane antiferromagnetic order (AFM). This in-plane FM canting was not previously reported. Such in-plane FM canting is impossible to theoretically interpret using the usual kagome Heisenberg-DM model and thus demands theoretical analysis. Second, the magnetization and its FC-ZFC splittings imply an unusually wide regime of short range order above the  long-range AFM transition $T_L$, with small but increasing density of magnetic moments across a temperature window $T_L<T<T_*$. 

We proceed to theoretically analyze this behavior in two steps. First (Sec.~III), we show that the known large Dzyaloshinskii-Moriya exchanges imply an additional secondary exchange which is a kagome lattice variant of Kitaev exchange. We analyze the effects of this magnetic exchange, as well the effects of a subdominant bond-Ising anisotropic exchange, and find (Sec.~IV) that these are sufficient to produce the  unusual in-plane FM canting out of (local or global) in-plane antichiral AFM order. 

Second, we propose and investigate a model of spin clusters with short ranged AFM order to describe the regime $T_L<T<T_*$. Since the kagome-Kitaev canting mechanism described above also applies to the short ranged AFM, such spin clusters would gain a small (canted) FM moment, enabling them to contribute to magnetization as FM spin clusters, with an unusually reduced magnetization. We show (Sec.~V) that a phenomenological model of such weak-FM spin clusters can capture the $M(B)$ field-dependent magnetization isotherms across the temperature window $T_L<T<T_*$. To analyze the role of ABH defects in the spin-cluster short range order regime, we perform exact diagonalization of a 12-site model of an ABH defect embedded in an antichiral ordered environment (Sec.~VI) and identify complementary `classical' and `quantum' mechanisms that can inhibit the formation of long range order.

The rest of the manuscript is organized as follows. In Sec.\ \ref{sec_exp}, we give the details of the experimental observations. In Sec.\ \ref{sec_antichiral}
we give the minimal model for the antichiral AFM order, containing Heisenberg and Dzyaloshinskii-Moriya terms, and derive the terms that then necessarily appear at higher order in the spin orbit coupling, including the kagome Kitaev terms. In Sec.\ \ref{sec_kitaev}, we give the details of the Kitaev term and derive the associated in-plane canting via a classical  $q=0$ analysis. Sec.\ \ref{sec_cluster} describes the phenomenology of the FM cluster model and its connection to the experimental observations.
Finally, in Sec.\ \ref{sec_ABH}, we describe the role of ABH disorder on the magnetic order and indicate the mechanisms of obtaining an extended regime of short-ranged order. The details of various calculations are further given in the Appendixes.

\section{\label{sec_exp}Experimental findings: in-plane ferromagnetic canting and extended regime of short ranged order}

\begin{figure}[t]
\includegraphics[width=0.45\textwidth]{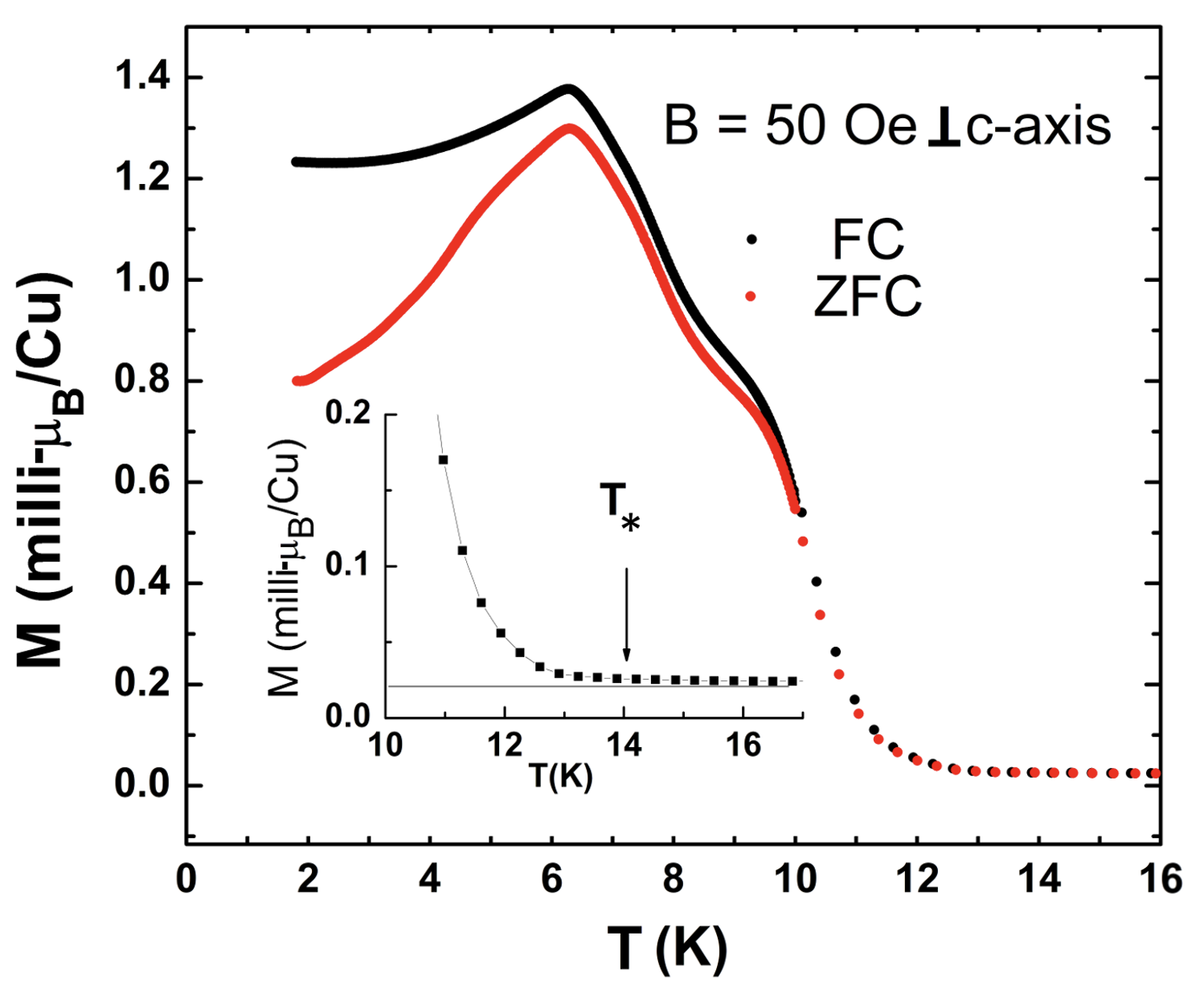}
\vspace{2.1cm}
\caption{\label{fig1} Temperature dependence of the magnetization, $M$, for the 58\% chlorine doped sample under FC and ZFC conditions (main panel) for a field of 50 Oe in the kagome plane (perpendicular to $c$). The hysteresis with strong FC-ZFC separation is visible below the $M(T)$ peak showing long range order below $ T_L \approx 6.5$ K.   Additionally, a slight separation between the two curves is visible from 10--12 K and slowly increases down to $T_L$, which suggests formation of ferromagnetic clusters across this temperature range $T_L < T < T_*$.  The inset is the same data on an expanded scale to demonstrate that the deviation from paramagnetic behavior sets in at $T_* \sim 14$ K. }
\end{figure}

\begin{figure}[t]
\includegraphics[width=0.35\textwidth]{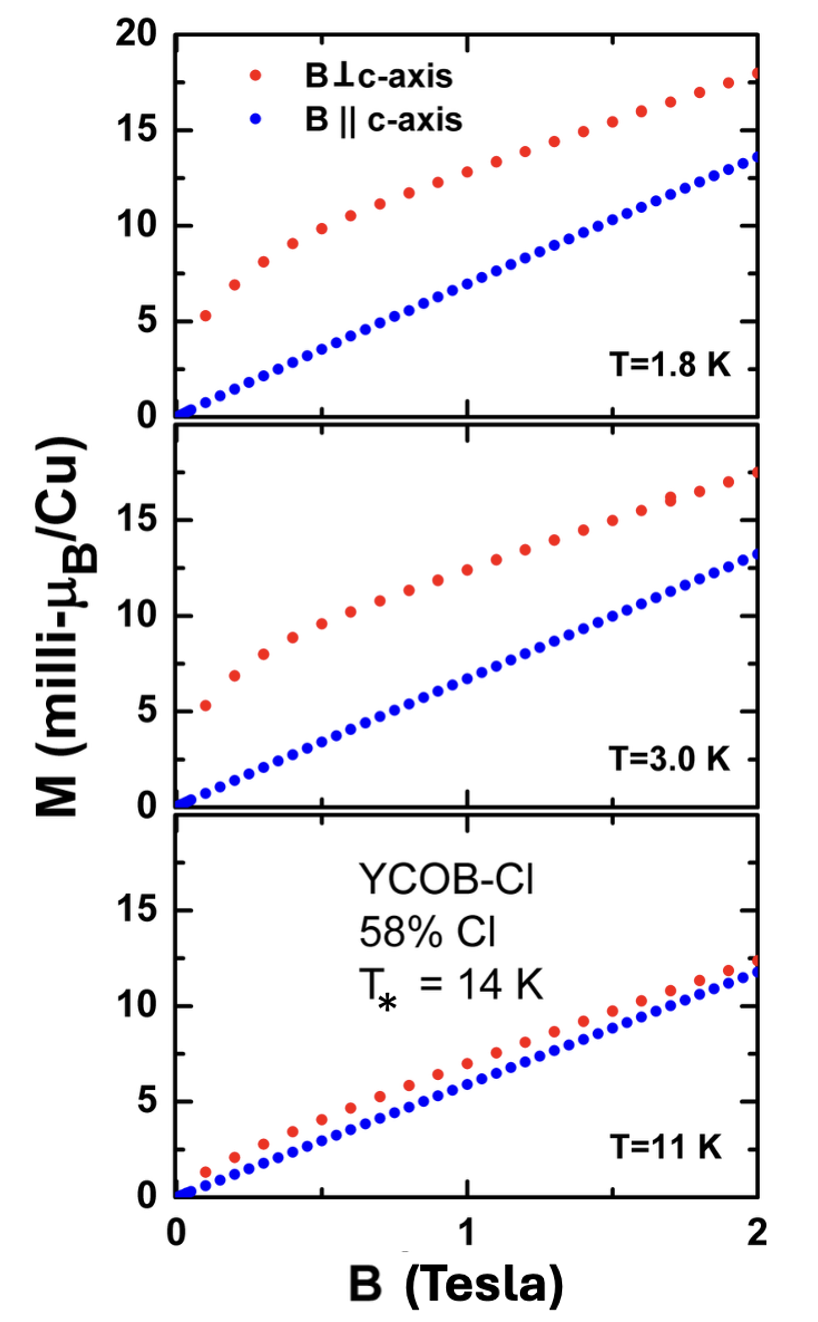}
\vspace{2.2 cm}
\caption{\label{fig2} Magnetization ($M$) isotherms (as a function of magnetic field $M(B)$ at fixed temperature $T$) here shown at three different temperatures below $T_*$. For B in the plane  ($\perp c$) (red dot-space-dot upper curves) a clear ferromagnetic response, suggesting some saturation at a fairly low field, is visible. The low field region has a larger slope compared to the high field linear response with a smaller slope. At low $T$ the ordered moment per site is of the order 0.01 $\mu_B$. As $T$ increases towards $T \rightarrow T_*$ the low field response shrinks and is absent above $T_*$. For B$\parallel$c-axis (blue dot-dot lower curves) no such ferromagnetic behavior is seen. Together, this suggests $T_L < T < T_*$ formation of spin clusters with in-plane FM moments, as shown below (Fig.~\ref{fig_spinclusters}).}
\end{figure}

\begin{figure}[t]
\includegraphics[width=0.45 \textwidth]{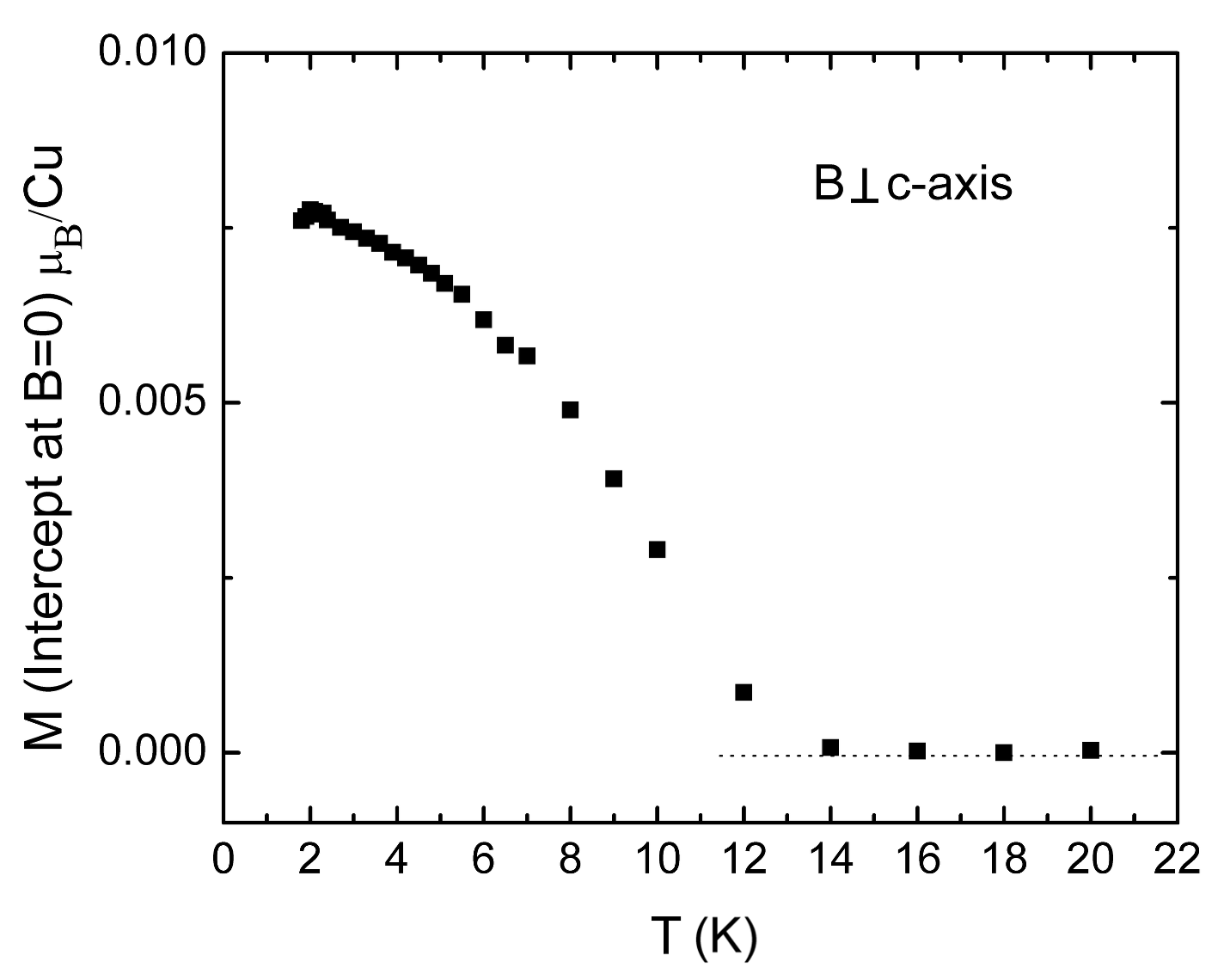}
\vspace{2.1cm}
\caption{\label{fig3} Implied in-plane ferromagnetic moment as obtained from an extrapolation of the high field (in-plane field  $\perp c$) linear behavior to obtain a $B=0$ intercept. The data implies a net ordered moment in the plane.}
\end{figure}

Here we present measurements on samples with $x=0.58$ and $y=0.1$.  The Cl concentration is large enough such that an ordered state is observed \cite{Xu2023}.   However, the magnetic order is quite unusual.
For this material (YCOB-Cl with x=0.58 and y=0.1) prior measurements \cite{Xu2023} of $ C(T)$, $\chi(T)$, and NMR all show weak magnetic order with some glassy aspects and with clear hysteresis below $T_\text{L} \approx 6.5 $ K.
This behavior can also be seen in the magnetization $M(T)$ shown in Fig.~1.

In Fig.\ 1 we show the magnetization as a function of temperature  measured on a sample with x=0.58 \% chlorine and y=0.1.  
With decreasing temperature there is at first a onset of short-ranged AFM order around $T_*\sim 14 $ K. 
Then a sharp change in slope occurs at the AFM transition of 6.5 K. 
The magnetization is shown
under both field cooled and zero field cooled conditions, with the field in the kagome plane  ($\perp c$).

The key new findings concern in-plane FM moments. 
The field cooled and zero field cooled curves gradually separate, indicating hysteretic behavior which is  associated with ferromagnetism. This FM component is observed to be strongly anisotropic occurring for fields in the plane.
The FC-ZFC separation is initially very small but opens up suddenly at the long-range order  temperature $T_L$$\approx$ 6.5 K.  We interpret this as the  onset of large well-ordered ferromagnetic regions which initially occupy a small volume fraction of the sample. These regions grow and overlap to form a bulk FM below 6.5 K.   This behavior is very different from what is expected for a pure AFM, which shows a cusp in the magnetic susceptibility at the Neel temperature. The in-plane FM canted moments are clearly visible.

In Fig.\ 2, we present the isotherms of the magnetization obtained at three different temperatures as shown.  
The isotherms for in-plane fields exhibit a dual slope response - with an initial rapid rise which fades away as the temperature is raised towards $T_*$. These isotherms clearly suggest a model  of ferromagnetically coupled spins that contribute to the rapid rise but soon attain a saturated state. Note that beyond saturation at higher fields there is  a separate contribution to the magnetization arising from the antiferromagnetic coupling of the spins.
Extrapolating the high field linear $M$ regime to low fields, we obtain the $B=0$ intercept of magnetization, which is plotted in  Fig.~3. The implied ordered moment lies completely in the plane. Its magnitude is small, below one percent, but increases as $T$ decreases from $T_*$.

This small FM moment coexists with the primarily antiferromagnetic order. Though neutron scattering is not available on this particular x=0.58, y=0.1 compound, there is good evidence that the AFM order here is the antichiral AFM order. 
Antichiral AFM (the $q=0$ 120$^\circ$ spin structure with negative vector chirality) has been unambiguously confirmed using neutron scattering in the original YCu$_3$(OH)$_6$Cl$_3$ undoped member of this material family  \cite{ZorkoPRB2019negative}.  It is also expected to describe the magnetism in the present $x=0.6$ $y=0.1$ compound \cite{Xu2023}, which has very similar specific heat and susceptibility behavior as YCu$_3$(OH)$_6$Cl$_3$. Both also have a transition at a similar temperature. Note that neither has a visible specific heat anomaly. While the susceptibility and specific heat, and the temperature scale of the AFM transition, are similar between YCu$_3$(OH)$_6$Cl$_3$ and the present compound, they are quite different in the other compounds of the family that exhibit $q=(1/3,1/3)$ order. The existence of the AFM transition in the present compound is also evident in NMR and ESR \cite{Xu2023}.
Below we will also show that a  positive-chirality  120$^\circ$ order is also ruled out by the experimental observation of in-plane canting.

\begin{figure}
    \centering
    \includegraphics[scale=0.12]{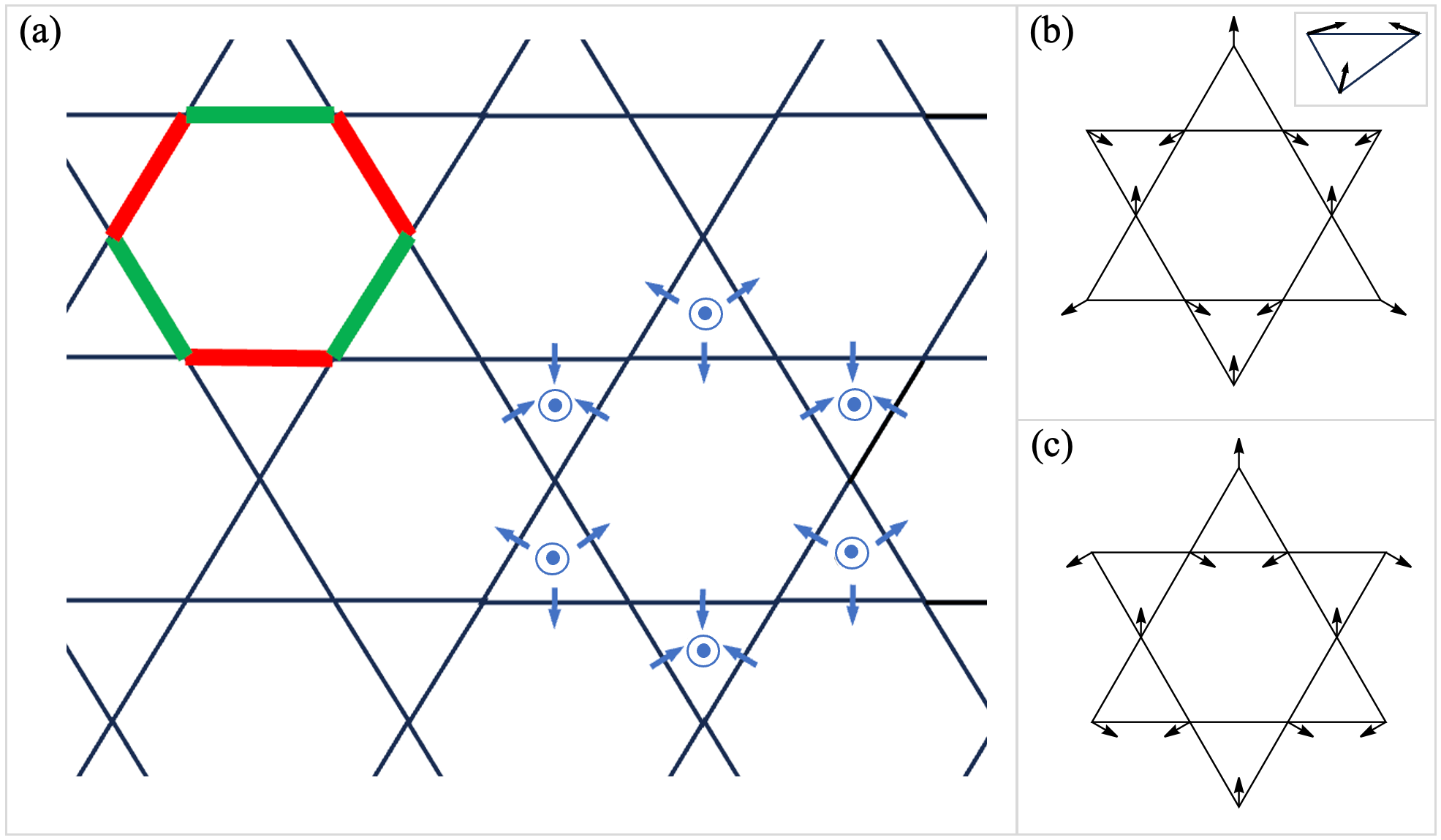}
    \vspace{2.2cm}
    \caption{Magnetic ordering in kagome lattice due to Heisenberg and DM interaction. (a) kagome network with dilute alternating-bond-hexagon (ABH) defects shown with thick red/green  bonds hosting stronger/weaker exchange couplings respectively. (ABH also appear with red/green interchanged, i.e.\ 2$\pi/6$ rotated.)  The blue arrows show in-plane DM vectors on each bond. The blue dots denote the out-of-plane DM vector on the surrounding bonds. To define the sign of DM vectors, bonds are oriented counterclockwise around each hexagon, or equivalently, clockwise around each triangle.  (b) Chiral magnetic order for $D_z>0$. Inset shows the ferromagnetic canting due to in-plane DM coupling. (c) Antichiral magnetic order for $D_z<0$. This is the magnetic order considered below and relevant to the material, which has no out-of-plane canting. }
    \label{fig_antichiral}
\end{figure}

\section{\label{sec_antichiral}Minimal spin model for antichiral AFM  order with in-plane canting}

\begin{figure*}
    \centering
    \subfigure[]{\includegraphics[scale=0.36]{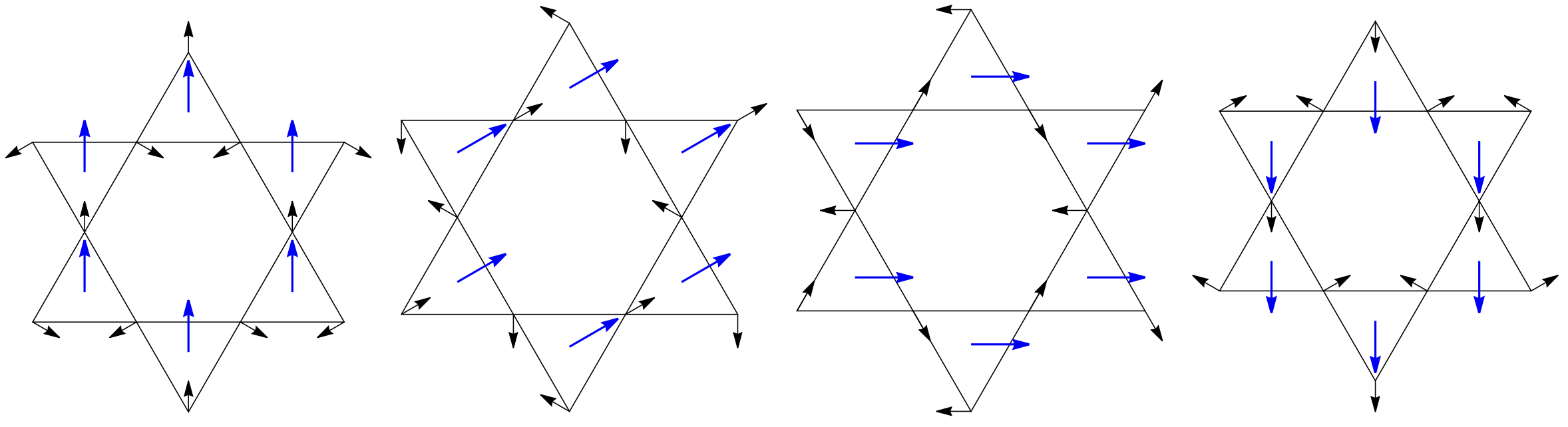 }}
    \subfigure[]{
    \includegraphics[scale=0.36]{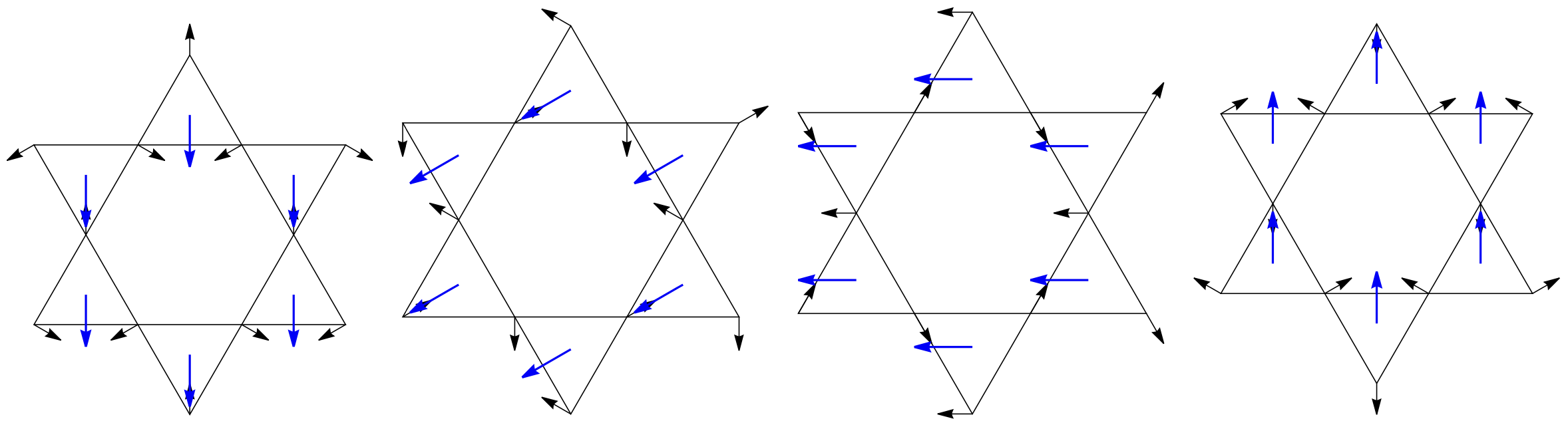} }  
    \vspace{2.3cm}
    \caption{
    Various AFM antichiral spin configurations (black) and associated FM canting (blue) arising by adding Kitaev interactions with  AFM (a) or FM (b) sign. The black arrows on the sites (kagome vertices) show the spins of the corresponding antichiral order. The blue arrow on the center of each triangle represents the net moment of the corresponding triangle. The canting due to AFM (a) and FM (b)  Kitaev  interaction differ  by the sign reversal of the canted moment (blue arrows). The four figures on each panel show four representative configurations (or domains) of the antichiral order (and the associated different configurations of the canting moment) given by $\phi_1=~\pi/2,~5\pi/6,~\pi,~3\pi/2$. Though at the classical level $\phi_1$ has an emergent $U(1)$ symmetry, which is moreover preserved by the Kitaev exchanges, we choose these values because we expect higher order quantum fluctuations to break the emergent $U(1)$ down to the discrete microscopic lattice symmetry of $2\pi/6$ rotations. 
    The four domains with rotated spin angle $\phi_1$ show the unusual feature of the canting (Eqn.~\ref{eqn:canted}): under rotations of the  ordered spin moments, the net canted moment $m$ rotates  oppositely.
    }
    \label{fig_canting}
\end{figure*}

In this section, we give a microscopic mechanism of how to obtain the spin interactions which we show in the next section can produce in-plane ferromagnetic canting.

The most dominant symmetry allowed spin exchange Hamiltonian on ideal single layer  kagome lattice can be written as~\cite{ElhajalPRB2002}
\begin{align}
    &H=J\sum_{\langle ij\rangle} S_i\cdot S_j+D_z \sum_{\langle ij\rangle }\left(S_i\times S_j\right)_z+\cdots,
    \label{eq_heis_dm}
\end{align}
where the first and the second term denote the isotropic Heisenberg and anisotropic DM interaction with DM vector being perpendicular to the kagome plane (see Fig. \ref{fig_antichiral}(a)), respectively,  and ``$\cdots$" denote other symmetry allowed couplings which will be discussed later. (We follow the convention of labeling the $c$-axis out-of-plane DM term by $D_z$.) 
Note that the  DM coupling is allowed due to the lack of inversion symmetry about the kagome bond center, and its strength is linearly proportional to the spin-orbit coupling  ($\lambda$), $|D_{z}|\propto \lambda J$~\cite{ElhajalPRB2002}. 
In this family of materials, the DM term is typically estimated to be $\mathcal{O}\left(10^{-2}\right)$ smaller than the Heisenberg interaction~\cite{lee2018}.
For the Heisenberg only term, the model is frustrated and favors a coplanar 120$^0$ AFM states on the kagome network at low temperatures via order by disorder. 
However, the DM term selects states with specific handedness among that subspace  via a finite temperature phase transition. For $D_z>0$, the ``chiral" structures are chosen which preserve the $C_3$ rotation symmetry of the kagome lattice about the midpoint of a triangle, see Fig. \ref{fig_antichiral}(b). The symmetry acts on both the spin and spatial (spin-orbit coupled) indices of the magnetic degrees of freedom.  For $D_z<0$, ``antichiral" structures are chosen which do not have any  physical $C_3$ rotation symmetry, see Fig. \ref{fig_antichiral}(c). In other words, if spin and space were unphysically decoupled, it would have rotation symmetries which act oppositely on spin and on space.

The consideration of the crystal field environment around the magnetic sites generically breaks the mirror symmetry about the pure  kagome plane,  and allows the  in-plane DM terms (perpendicular to the kagome bonds, see Fig. \ref{fig_antichiral}(a)) in the above Hamiltonian. Although they also appear at the linear order in spin-orbit coupling, they are much weaker than the out-of-plane DM term (because they arise from inter-plane effects), hence their effects on the chiral and antichiral states can be computed perturbatively. For the chiral states, the in-plane terms generate an out-of-plane canting of the spins leading to a net FM moment perpendicular to the kagome plane.  
The relevant material showing no out-of-plane magnetization then rules out the possibility of the chiral order.  
The antichiral states, however, shows no out-of-plane canting with the in-plane DM terms. This can be clearly seen from a symmetry argument: the in-plane DM vectors transform oppositely under $C_3$ rotations compared to the antichiral spin order. Thus there is no energy gain due to the DM term for canting and the cost due to the Heisenberg exchange forbids it.

Focusing now on the antichiral state, we look for a mechanism for in-plane canting as observed in experiment. In-plane DM terms cannot generate such canting because the DM energy remains insensitive to any in-plane FM moment. (This is true even for the chiral state.)
Therefore, we must look for other symmetry allowed couplings that must be added to Eq. \ref{eq_heis_dm}. 
%
In presence of  spin-orbit coupling, such exchanges can be obtained in a standard manner \cite{moriya_anisotropic_1960}. Given the DM vector on a bond connecting site $i$ and $j$ being ${ \hat{D}}_{ij}$,  the symmetric easy axis anisotropic interaction is given by,
\begin{align}
    H_\text{soc, sym}= \frac{1}{J}\left({ S_i}\cdot{\hat{D}_{ij}}\right) \left( { S_j}\cdot { \hat{D}_{ji}}\right).    
\end{align}
Clearly, such term varies quadratically with the spin-orbit coupling and remains weak compared to other exchanges described previously.
Since the net DM vector in kagome is a linear combination of out-of-plane (perpendicular to kagome plane) and in-plane (perpendicular to the kagome bond) component (see Fig. \ref{fig_antichiral}(a)), it leads to following decomposition of the anisotropic exchange interaction:
\begin{align}
    H_\text{soc, sym}=\alpha_1 H_{c\text{-Ising} }+\alpha_2 H_\text{Kitaev}+\alpha_3                 H_\text{cross},
\end{align}
where $\alpha_i$s are constant coefficients which depend on the ratio of the in-plane and out-of-plane DM component. In the above decomposition, $H_{c\text{-Ising}}\propto \sum_{\langle ij\rangle}S^c_iS^c_j$ denotes the Ising interaction with easy axis perpendicular to the kagome plane ($c-$axis). 
$H_\text{Kitaev}$ is a bond dependent Ising interaction~\cite{kimchi_kitaev_2014}: 
\begin{align}
   & H_\text{Kitaev}=J_\text{Kitaev}\sum_{\langle ij\rangle_\mu}S_i^\mu S_j^\mu~~(\mu=X,Y,Z),
\end{align}
where the Kitaev axes $X,~ Y$ and $Z$ correspond to each of the three bond orientations on the kagome lattice. Explicit expressions are given in the Appendix \ref{appen_kitaev_ising}. $H_\text{cross}\propto \sum_{{\langle ij\rangle}_\mu}\left(S_i^cS_j^\mu+S_j^cS_i^\mu\right )$ denotes the cross term between Kitaev and $c$-axis Ising term. We note that $c-$axis Ising term cannot generate the in-plane ferromagnetic canting, whereas $H_\text{cross}$ gives the in-plane canting only to the subleading order. Therefore, the dominant in-plane canting arises from the Kitaev term which we explicitly compute in the following section. 

We note that in other kagome systems there could be other symmetry allowed bond dependent Ising interactions with easy axes lying in the plane, that also can potentially give rise to the in-plane FM canting. One of such example is the bond-Ising interaction. (The bonds are in-plane so this exchange is orthogonal to the c-axis Ising exchange.) However, in the present case the Kitaev interaction gives the most dominant contribution as it directly appears from the microscopic spin-orbit coupling. For completeness we discuss canting effects from the bond-Ising term in the Appendix \ref{appen_kitaev_ising}. We now proceed to focus on the Kitaev-type interaction to show that it does indeed produce canting.

\section{\label{sec_kitaev}In-plane FM canting via kagome Kitaev exchanges}

Expanding  the classical Kitaev energy around the antichiral state perturbatively within a single triangle ($q=0$) analysis appropriate for this $q=0$ order, we obtain:
\begin{align}
    \frac{H_\text{Kitaev}}{J_\text{Kitaev}}
   &\approx -\frac{1}{2}+\frac{1}{6}\left(a\cos(2\phi_1)+b\sin(2\phi_1)\right)+\mathcal{O}(a^2,b^2,ab),
   \label{eq_classical energy_1}
\end{align}
where $a$ and $b$ denote the in-plane deviations of the spin configuration from antichiral order: 
\begin{align}
    &\phi_2-\phi_3=\frac{2\pi}{3}+\frac{a}{\sqrt{3}}~~, ~~\phi_2+\phi_3-2\phi_1=2\pi+b,
    \label{eq_expansion around antichiral_ising}
\end{align}
and, $\phi_1,~\phi_2,~\phi_3$ are the angle variables measured from the $x-$axis (see Fig. \ref{fig_reference} in the Appendix), describing any in-plane spin configuration with the indices defined in the anticlockwise manner (As an illustrative example, see the triangle formed by sites 7, 1 and 2 in bottom right inset of Fig. \ref{fig_quantum analysis_hexagon_spin expectation_magnetic field}). 

The Kitaev interaction explicitly breaks the continuous $U(1)$ spin rotation symmetry.  
However, the leading  term in the energy is a $\phi_1$ independent constant, which implies although the Kitaev term breaks the $U(1)$ explicitly, there is still an emergent $U(1)$ invariance in the antichiral subspace at least at the classical level.

The classical Kitaev energy is minimized by the following spin configurations:
\begin{align}
    &J_\text{Kitaev}>0 ~\text{(AFM)}~: ~f_1= 2\phi_1+\pi, \\ &J_\text{Kitaev}<0 ~\text{(FM)}~: ~f_1= 2\phi_1,
\end{align}
where, $f_1=\tan^{-1}(b/a)$. Clearly, the canting configuration is tied to the overall AFM order parameter spin angle $\phi_1$. 
The net canted moment on the triangle is given by
\begin{align}
    m=\Bigg\{\begin{array}{c}
    \frac{1}{2}\left\{\cos\phi_1~,~-\sin\phi_1\right\}~,~~\text{for FM Kitaev.} \\
    \frac{1}{2}\left\{-\cos\phi_1~,~\sin\phi_1\right\}~,~~\text{for AFM Kitaev.}
    \end{array}\label{eqn:canted}
\end{align}
The subdominant bond-Ising interactions give the same canting tendencies for AFM and FM Ising respectively (see the Appendix for details).

The canting produces a net moment from any region of uniform antichiral ordering with a uniform choice of the global symmetry breaking angle $\phi_1$. This is visually shown in Fig.~\ref{fig_canting} which plots the local canted moment for various antichiral spin configurations. Figure \ref{fig_canting} also shows the unusual feature of Eqn.~\ref{eqn:canted}: under rotations of the  ordered spin moments on each site, the net canted moment $m$ rotates \textit{oppositely}.

Since the canting mechanism is local, such canting can also occur locally even if the true long-range ordering is absent and  there is only a short-ranged antichiral state present over a finite cluster, with fluctuating spin configuration and fluctuating net moment.

For completeness we also computed the possibility of local in-plane canting arising from the ABH disorder. However,  due to the $C_3$ rotation symmetry of the ABH, the resultant FM moment of the triangles forms a vortex configuration with zero average moment per hexagon; see  Appendix \ref{appen_ABH} for further details.

\section{\label{sec_cluster}Phenomenology of short ranged FM order via spin clusters}

\begin{figure}[t]
\includegraphics[width=0.45\textwidth]{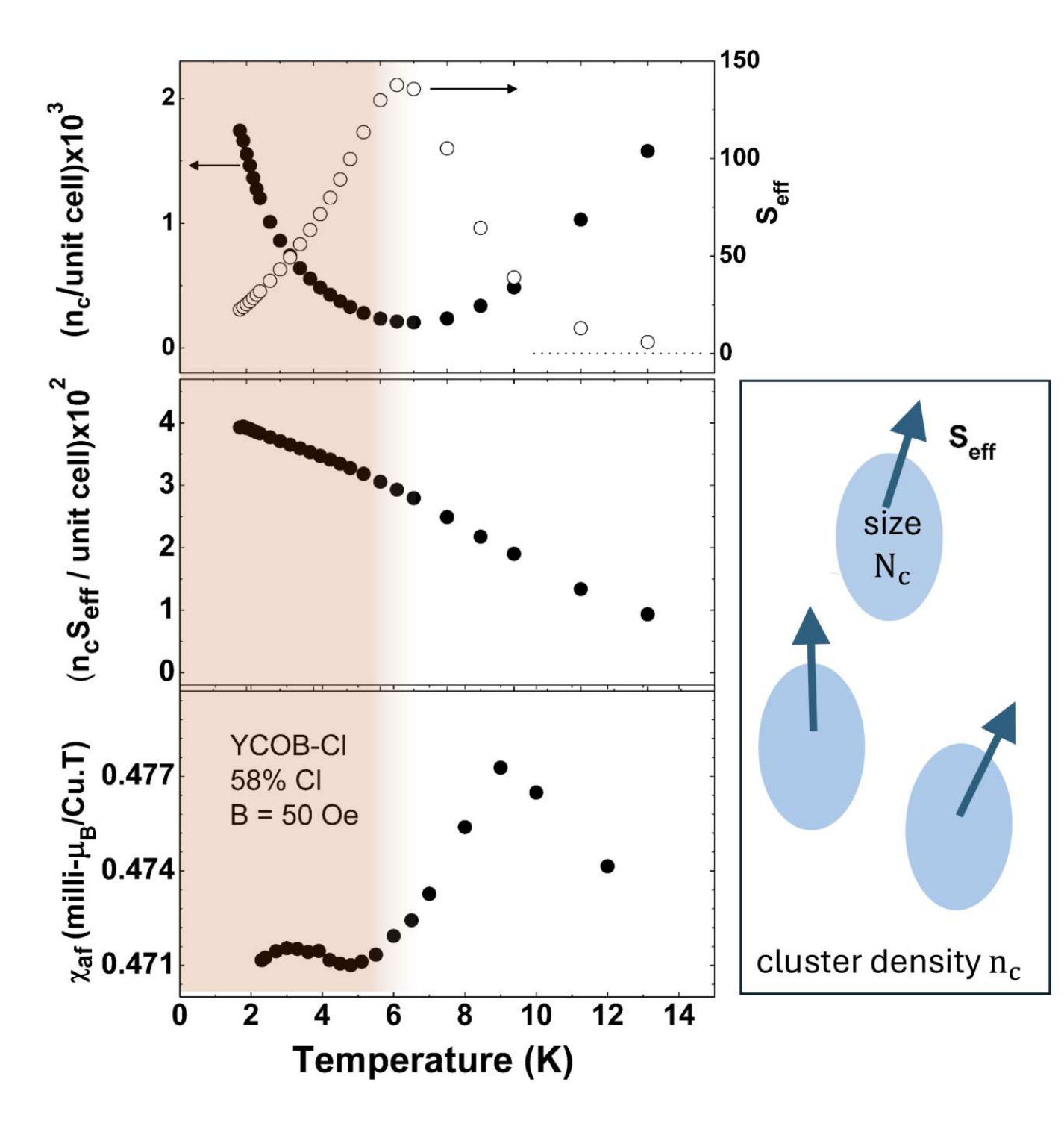}
\vspace{2.2cm}
\caption{Phenomenological spin-cluster description of  the short-ranged ordered regime. Right panel: Schematic diagram for the spin-cluster model. Cluster size $N_c$  is related to the effective in-plane moment of the cluster $S_\text{eff}$ by $N_c\sim S_\text{eff}/\alpha_m$, with $\alpha_m$ being the small fraction of moment per site participating in FM canting.   Left: spin-cluster formation as obtained through fits to $M(B)$ isotherms as in Fig.~\ref{fig2} using Eqn.~\ref{eq:Mfit}. The fit to the measured function $M(B)$ was performed at each $T$, to extract the high-field linear coefficient $\chi_{AF}$ and the nonlinear Langevin spin-cluster magnetization described by the cluster density $n_c$ and effective moment $S_\text{eff}$. Top panel shows  $n_c$ (closed circles) and $S_\text{eff}$ (open circles). Their temperature dependence shows cluster merging via increasing $S_\text{eff}$ and decreasing  $n_c$ across the regime $T_L<T<T_*$. The onset of long range magnetic order at $T_L\approx 6.5$K is visible  as saturation of  $S_\text{eff}$ and   $n_c$. For $T<T_L$ (orange shaded region) the cluster model breaks down, as self consistently shown by the opposite $T$ behavior of $S_\text{eff}$ and   $n_c$. This shaded region is shown for completeness but cannot be interpreted in terms of clusters. Middle panel shows the product of $n_c$ and $S_\text{eff}$ which varies smoothly throughout the measured temperature range as discussed in the text. Bottom panel shows  $\chi_\text{AF}$ extracted from the linear high-$B$ behavior at each $T$.  
The shift of the $\chi_\text{AF}$  peak from the physical $T_L$ value is due to this high-field limitation for $\chi_\text{AF}$ in the $M(B)$ fits; also note the small range of $\chi_\text{AF}$ variation in the y-axis scale. Sections V and VI are concerned with the cluster growth regime $T_L<T<T_*$.}
\label{fig_spinclusters}
\end{figure}

Having discussed the origin of the small ferromagnetism, we now focus on the second part of the observed puzzle of whether there is extended regime of short-ranged order below $T_*$ before the onset of true long-ranged magnetism. We answer this question in two parts. In this section we phenomenologically describe a spin-cluster model, and show that the short-ranged clusters along with their  in-plane moments due to Kitaev interaction describes the experimental observations quite well. In the next section we introduce a  microscopic model for the enhanced short range order by studying the interplay of ABH disorder and antichiral AFM order in a 12-site ABH toy model.

The phenomenological spin-cluster picture described here is  reminiscent of Griffiths-phase effects which can occur in the vicinity of phase transitions in disordered systems \cite{Vojta2013}. 
However, here the long range order is primarily AFM antichiral order, while the signatures of spin clusters arise due to their weak FM canting. The weak FM canting of short ranged AFM regions, producing small moment spin clusters, is unconventional. An additional difference concerns the microscopic source of disorder. As discussed in the next section, the localized ABH lattice defects, rather than rare regions of long wavelength disorder, are sufficient for understanding the formation of short range order clusters. This microscopic disorder analysis also suggests that the spin-cluster effects should here be highly visible above the ordering transition, as is indeed seen in the fitting results we describe below.

We now proceed with a phenomenological analysis of spin-cluster effects. 
Phenomenologically, we attribute the $T_*$ temperature to the onset of a short-ranged correlated phase, where the magnetism is controlled via independent finite spin clusters. Such clusters can grow and gain  a small in-plane ferromagnetic moment as we lower the temperature. The  magnetization of the clusters can then be written as 
\begin{align}
M(B;T) &= n_c(T)\  g \mu_B S_\text{eff}(T)\  L\left(\frac{g\mu_B S_\text{eff}(T)}{k_BT}B\right) 
\nonumber \\
&+ \chi_\text{AF}(T) \ B .
\label{eq:Mfit}
\end{align}

In this equation the magnetic response is split into two components.  The first term corresponds to the response of the free spin clusters, which contribute a nonlinear Langevin function response at small field $B$. The second describes the response of the AFM which gives a small-slope linear response at large $B$. Since this function describes the magnetization as a function of field $B$ at any temperature $T$, we write the explicit temperature dependence for the temperature-dependent parameters $n_c(T), S_\text{eff}(T)$, and $\chi_\text{AF}(T)$. As we describe below, these are the parameters we fit to $M(B)$ at each temperature $T$.

 The spin-cluster magnetization is for simplicity here captured with just two free parameters. Here
$S_\text{eff}(T)$ and $n_c(T)$  describe the effective net magnetic moment for each cluster, and the density of the clusters (number of clusters per number of unit cells), respectively. These are generically temperature dependent quantities. Additionally the fixed parameter $g=2$ denotes the Land\'e g-factor for coupling external magnetic field and FM clusters. 
In the above equation, the free spin-cluster magnetization is approximated by the Langevin function $L(x)=\coth(x)-1/x$ which gives the magnetization of a free classical spin as a function of the dimensionless combination $x$. This ratio $x$ is the magnetic energy $g\mu_BS_\text{eff} B$ divided by temperature $k_B T$. The Langevin function is the large spin (large $S_\text{eff}$) limit of the Brillouin function which gives the magnetization of a quantum spin-$S$ moment. This limit amounts to a classical description of the net FM cluster moment. Finally, $L(x)$ comes multiplied by a coefficient, given by the product of the density of spin clusters $n_c$ times the magnetic moment per cluster $g \mu_B S_\text{eff}(T)$.

We fit the measured $M(B)$ isotherms as in Fig.~\ref{fig2} using Eqn.~\ref{eq:Mfit}. The fit to the measured function $M(B)$ was performed at each $T$, to extract the high-field linear coefficient $\chi_{AF}$ and the nonlinear Langevin spin-cluster magnetization described by the cluster density $n_c$ and effective moment $S_\text{eff}$. This fitting procedure is robust, as can already be seen visually from inspection of Fig.~\ref{fig2}. The high field region is linear and provides the AFM linear contribution $\chi_{AF}$ for the fit. The nonlinear behavior at small $B$ is fitted to the Langevin function $L(x)$ with just two parameters, cluster density $n_c$ and effective moment $S_\text{eff}$, which together just act as a magnetic field scale $x$ in  $L(x)$ and as an overall magnetization coefficient multiplying $L(x)$. Since $S_\text{eff}$ is necessarily a dimensionless number quite a bit larger than 1,   as indeed seen in the fit, the Langevin function saturates for quite small fields, producing a nonlinear small $B$ contribution that can be distinguished from the large-$B$ slope of $\chi_{AF}$.

In Fig.~\ref{fig_spinclusters}, we plot the parameters $n_c$ (closed circles), and $S_\text{eff}$ (open circles) obtained from fits to the $M$ vs $B$ isotherms at each temperature. 
The interpretation of the resulting fit in terms of spin-cluster formation in the temperature regime $T_L<T<T_*$ is evident from the behavior of the fitted parameters in this regime. 
We see that $S_\text{eff}$ rises rapidly below 14 K, with the number of clusters decreasing in a concomitant manner. Their product (middle panel) is a smooth function of $T$. Interestingly, $\chi_\text{AF}$ shows a peak (bottom panel) as expected for an AFM transition, though the peak position extracted from this few-parameter $M(B)$ fit, which effectively restricts $\chi_\text{AF}$ to only sample high fields, is thus shifted from the true (low-temperature, low-field) AFM transition.

We can further interpret these fit parameters.
The free spin clusters giving a net magnetization have a typical size which depends on temperature. We estimate this cluster size $N_c$ as scaling with its total moment $S_\text{eff}$ by $N_c \sim S_\text{eff} / \alpha_m$, where the temperature-independent parameter $\alpha_m$ is the fraction of the moment per site that contributes to the FM net magnetization. 
The estimated cluster size $N_c = S_\text{eff} / \alpha_m$ gives the number of sites participating in the cluster. This should be distinguished from the density of clusters, $n_c$, which measures the number of clusters in the system. 

Interestingly, the product $n_c N_c \alpha_m =  n_c S_\text{eff}$ shows a strikingly simple temperature dependence (Fig.~\ref{fig_spinclusters} middle panel). Above  $T_L$, we interpret $f_c = n_c N_c$ as the fraction of all magnetic sites that participate in the clusters.   Long range magnetic order at $T_L$ is triggered when $f_c$ becomes of order $ \approx 1$. 
At 7 K we find that $n_c S_\text{eff} \approx 0.03$, implying $\alpha_m \approx 0.03$, which provides an estimate of cluster size: each cluster moment $\mu_B$ arises from the combination of about a hundred Cu sites.

For completeness we also perform the same fit for $T<T_L$, where the spin-cluster picture should break down. Indeed we find that the fit shows this breakdown self consistently. At $T_L$ the temperature derivative of $n_c$   and $S_\text{eff}$  change sign, signifying that the cluster merging process has completed.  For $T<T_L$ the fit parameters give  decreasing $S_\text{eff}$ which rules out the cluster picture. Instead, this can be interpreted to correspond to the decreasing correlation length $\xi$,  as $N_c = S_\text{eff} / \alpha_m \sim \xi^2$. Correspondingly the product $f_c = n_c N_c$ rises above 1, interpreted as a merged cluster percolating across the entire system, with an increasing FM moment per site corresponding to the long ranged FM order. Above $T_L$ the cluster picture is physical, and for the remainder of this manuscript we restrict ourselves to this temperature regime $T_L < T < T_*$ to study the short ranged order and possible microscopic mechanisms for the spin clusters with in-plane moments.

\section{\label{sec_ABH} Extended regime of short-ranged order: Role of Alternating-Bond-Hexagon disorder}

In this section, we will look for a mechanism of obtaining the spin clusters based on the ABH disorder present in the current material. In Fig. \ref{fig_antichiral}(a), the ABH disorder is depicted  with the red and green showing bonds with stronger and weaker spin exchange couplings. As described earlier, such alternation of coupling strength arises due to the random substitution of Br site by OH both above and below the kagome hexagons, hence the disorder density  is controlled by the parameter $y$, which is 0.1 for the the current sample.

\begin{figure}
    \centering
    \includegraphics[scale=0.266]{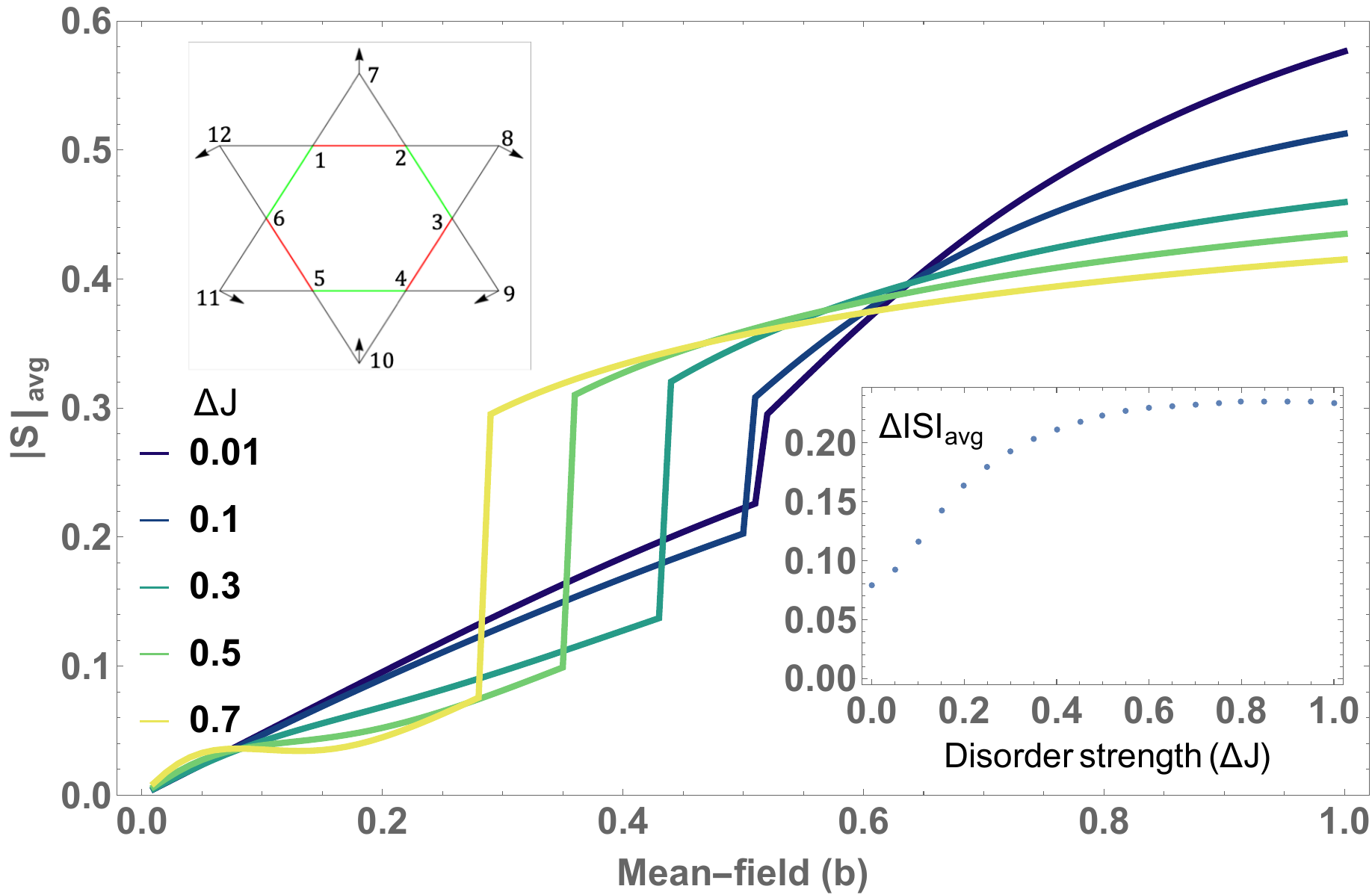}
    \vspace{2.2cm}
    \caption{ Exact diagonalization of 12-spin Hamiltonian on a star of David with ABH: Inset on the top left shows the star of David with ABH and mean-field assigned on outer vertices consistent with antichiral pattern. Red and green lines show strong and weak bonds with strength $J+\Delta J$ and $J-\Delta J$ (assuming $\Delta J$ positive), respectively. The black arrows on the outer vertices represent the effective  mean-field acting on those sites due to the short-ranged antichiral AFM order on its surrounding.
    In the main figure, average norm   of the ground state expectation values of the spins on the ABH, $|S|_\text{avg}=\frac{1}{6}\sum_{i=1}^6|\langle {\bf S_i}\rangle|$, is shown as a function of antichiral order mean-field. Each curve of the plot  are obtained by setting a fixed value of disorder strength $\Delta J$. The maximum of the physical mean-field considered is  $2J\langle |S|_\text{avg}\rangle<1$. 
    The sharp jump in $|S|_\text{avg}$ represents the location of the level crossing. The inset of the bottom right shows the variation of the change of $|S|_\text{avg}$ at the level crossing with disorder strength. To enable comparison with classical analysis here $S$ is given in units of $\hbar/2$.}
    \label{fig_quantum analysis_hexagon_spin expectation_magnetic field}
\end{figure}

To understand the role of ABH disorder we consider a fully quantum  12 site model consisting of a star of David surrounding an ABH, and diagonalize it exactly. Focusing a single ABH is justified if we assume  a dilute density of ABH  in the system. Every ABH is then locally surrounded by the pristine kagome which we further assume hosts an at least short-ranged antichiral order. 
We incorporate this surrounding short-ranged order in the model by applying a mean-field $b$ on the outer vertices of the star of David containing the ABH consistent with the surrounding antichiral pattern (see bottom right inset of Fig. \ref{fig_quantum analysis_hexagon_spin expectation_magnetic field}). This leads to the following 12-site Hamiltonian:
\begin{align}
    H = &H_\text{Heis}+\sum_{i=7}^{12} {\bf b}_i\cdot { S_i}\nonumber 
    +\Delta J H_{\Delta J}
    \\
    H_{\Delta J} &= \left(\sum_{\langle ij\rangle\in \text{strong}} S_i \cdot S_j -\sum_{\langle ij\rangle\in\text{weak}} S_i \cdot S_j  \right)
\end{align}
where 
$H_\text{Heis}$ is the isotropic translation invariant Heisenberg term  
$H_\text{Heis}=J\sum_{\langle ij\rangle} S_i\cdot S_j$, and 
$H_{\Delta J}$ is  Heisenberg exchange on strong minus weak bonds on the central hexagon.
In this section we use the unit convention $\hbar/2=1$ so that the spin operators $S$ are Pauli matrices, which enables a more direct comparison between quantum effects and classical energies of unit vector spins. 
The disorder strength is measured by the energy parameter $\Delta J$. This can be interpreted as a dimensionless $\Delta$ times $J$; we use the product for clarity. The short ranged order is expressed using 
${\bf b_i}$ which are the mean-fields  assigned on the outer vertices of the star of David  as shown on the bottom right inset of Fig. \ref{fig_quantum analysis_hexagon_spin expectation_magnetic field}. We assume their magnitude remains same for all the vertices, $|{\bf b_i}|=b$. 
We then investigate the consequence of such mean field on the spins on the  ABH. In Fig. \ref{fig_quantum analysis_hexagon_spin expectation_magnetic field}, we show the average magnitude of the spin expectation vector on the hexagon, $|S|_\text{avg}=\frac{1}{6}\sum_{i=1}^6|\langle {S_i}\rangle|$ as a function of  mean-field. There are two distinct effects observed in the 12-site quantum model, associated with complementary mechanisms, which we denote as ``classical" and ``quantum".

\begin{figure}
    \centering
    \includegraphics[scale=0.268]{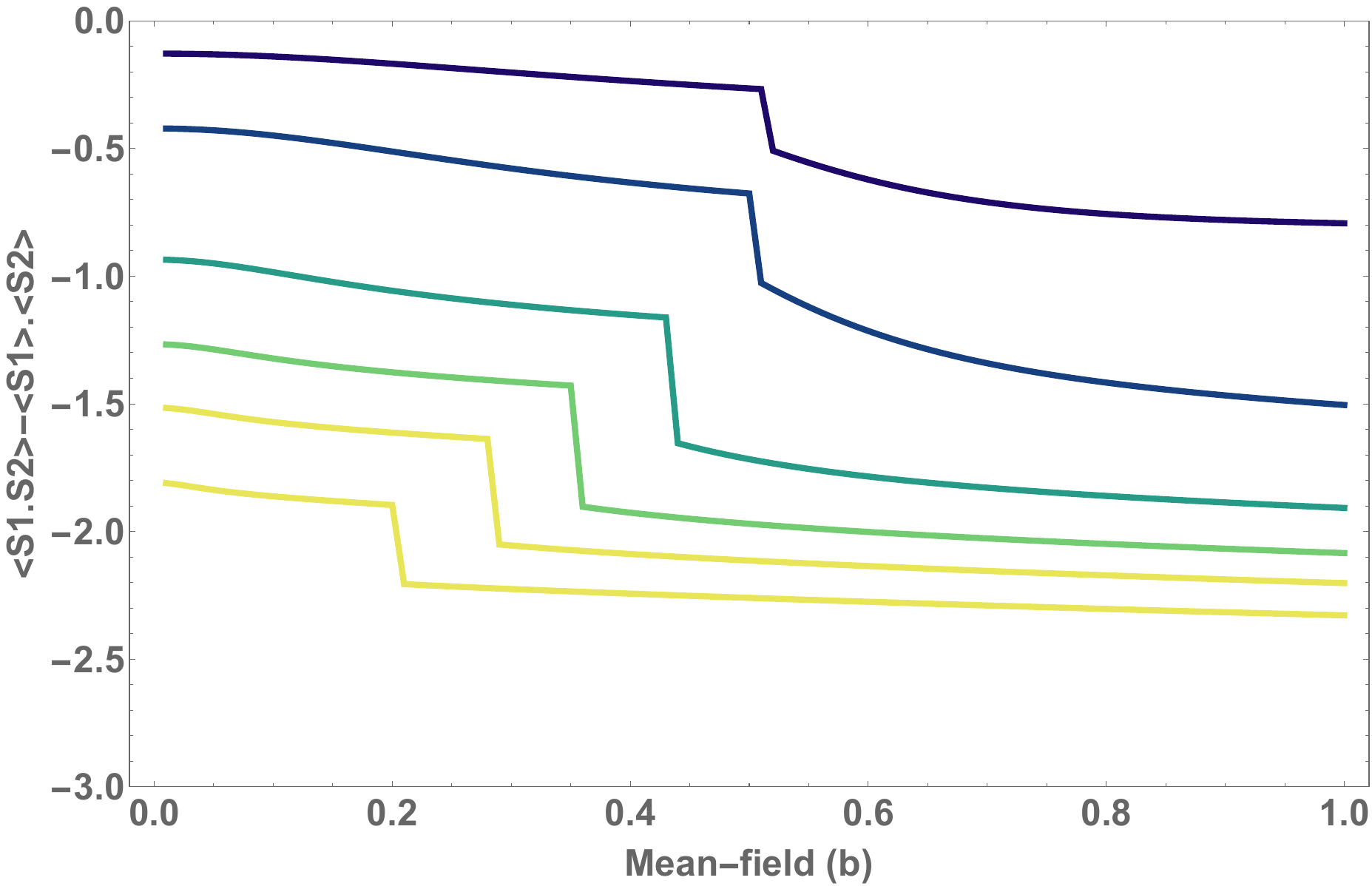}
    \vspace{2.2cm}
    \caption{Quantum fluctuations of bond energy on strong  bonds: this measure of the fluctuations of bond energy gives  -3 and 1 for an ideal singlet bond and $S^z=0$ triplet bond, respectively, in units of $\hbar/2$. With increasing  disorder strength the strong bond gains more of a singlet  character.    }
    \label{fig_quantum analysis_fluctuation_st}
\end{figure}

Classical mechanism: relieved classical frustration. 
The kagome lattice Heisenberg model is geometrically frustrated, with various possibilities of competing orders and frustration due to the competing interactions. The frustration can be seen already for classical Heisenberg interactions. Beyond the Heisenberg exchanges, the small DM coupling in addition to the Heisenberg interaction favors the antichiral states. This partially relieves the frustration. But the antichiral state is still geometrically frustrated, akin to other 120 degree states on lattices with triangles.

The ABH, incorporating the stronger and weaker couplings, partially relieves the frustration locally near the defect. It modifies the classical energy (per bond) of the classical antichiral order from $J/2$ to $J/2 - (\Delta J)^2/6J$. (See Appendix \ref{appen_ABH} for details.) Such energy modification is associated with the local FM canting of the spins on the disordered triangles, indicating further reduction of frustration. This local canting arising from ABH gives no net coarse-grained FM moment, but rather a larger-scale AFM pattern  (see Appendixes for details). The local reduction of frustration around the ABH can potentially help to set in short-ranged ordering on the ABH more easily than the rest of the pristine kagome, leading to the cluster model which has been introduced phenomenologically in the earlier sections. 

This effect is also seen when including quantum effects in exactly diagonalizing the 12-site ABH model: at larger $b/J > 0.2$, (sufficiently away from the classical ordered limit, $b/J < 0.5$) the classical spin moment $|S|_\text{avg}$ is greatly enhanced by $\Delta J$, with a positive jump arising with a level crossing with increasing $\Delta J$ (Fig. \ref{fig_quantum analysis_hexagon_spin expectation_magnetic field}). Phenomenologically, this enhanced $|S|_\text{avg}$ can be interpreted as a local enhancement of short-ranged order near each ABH. An interesting corollary of the above effect can be the pinning of density of the short-ranged cluster to that of the ABH, which may be assessed in future experiments that can study the regime of short ranged clusters while characterizing the ABH defects.

Quantum mechanism: enhanced singlet formation. 
Now considering quantum effects in full, it is clear that ABH supports singlet formation on the strong bonds, in addition to any other effects. Such singlet formation can successfully compete against classical orders: even the most naive valence bond solid states which consist of singlets on strong bonds and free spins elsewhere on ABH enjoy an energy per bond of $J/2 - \Delta J/2$. 

This effect is seen in the 12-site ABH model: at small $b/J \lesssim 0.2$ 
the classical spin moment $|S|_\text{avg}$ is suppressed by $\Delta J$ (Fig. \ref{fig_quantum analysis_hexagon_spin expectation_magnetic field}), and at any $b$ the strong bonds support increasing quantum fluctuations in the bond energy ($\langle S_i \cdot S_j \rangle - \langle S_i \rangle \cdot \langle S_j \rangle$) with $\Delta J$ (Fig. \ref{fig_quantum analysis_fluctuation_st}). Phenomenologically, this enhanced quantum fluctuation can result in suppression of long-range classical order and enhanced regime of short-ranged order arising from ABH. Such effects are especially quite significant near the $T_* \sim 14$ K, where the short-ranged order is weak enough. At much lower temperatures, the mean field due to the short-ranged domains enhances, leading to long-range ordered phase at $T_L \sim 6.5$ K. Clearly, the quantum effect of ABH can be thought of as stretching the domain of short-ranged ordering in the material.

\section{Discussion}

In this work we presented experimental puzzles in the magnetism of a kagome compound and suggested theoretical models to describe the puzzles. 
The experimentally observed  in-plane anisotropy of the FM canting and the extended regime of short-ranged order cannot be described by the usual geometric frustration or magnetic impurities on the kagome lattice. 
We theoretically approach this puzzle by considering each ingredient separately. To account for in-plane ferromagnetism we propose  additional unconventional interactions which are a kagome variant of the Kitaev exchanges. We show these are symmetry allowed and moreover microscopically expected due to the presence of already known DM interactions. We show that in-plane ferromagnetism is produced by such small Kitaev-like bond dependent Ising interactions. 
Given this theoretical mechanism for in-plane canting, we propose a phenomenological spin-cluster model to describe the unusual magnetism observed in the intermediate temperature regime of $T_L<T<T_*$. 
To identify a possible microscopic stabilization mechanism of the spin clusters and the associated expanded short ranged order regime, we then turn to the ABH defects. Using both quantum and classical analysis  we show that the ABH disorder stretches the stability of the short-ranged order  via both nucleation of the local order parameter around defects as well as by enhancing the quantum fluctuations that suppress a classically ordered state. The ABH disorder and resulting spin clusters, together with the weak but consequential kagome-Kitaev exchanges, can thus account for the unusual puzzle seen in the magnetism of the ordered compound. 
These theoretical results would apply to other kagome systems with similar magnetic signatures and show how in-plane canting can arise and how weak-FM spin clusters can be stabilized by disorder.

The identification of the salient experimental features of a quantum spin liquid state is an ongoing effort.  In the above we have presented comprehensive results on a relatively new system, YCu$_3$(OH)$_6$[(Cl$_x$Br$_{(1-x)}$)$_{3-y}$(OH)$_y$], in which replacing Br by sufficient Cl (i.e. tuning x) a magnetic state can be tuned in.  This substitution as well as the presence of the (OH)$^-$ ions necessarily introduces disorder. In the regime of disordered ABH density where there is an AFM ordered state, we found evidence for the formation of weak-FM clusters. We developed a theory for the origin of these clusters in the present setting, via disorder and Kitaev interactions.

Going beyond the magnetically ordered case towards a possible nearby Dirac QSL phase, we anticipate a complementary mechanism for generating FM moments from certain 120$^0$ AFMs.
For a Dirac QSL, chiral time reversal symmetry breaking with FM spin clusters generates a mass term for fermion bilinears, leading monopoles to proliferate and destroy the QSL; on the kagome lattice, the expected magnetic order is then 120$^0$ AFM \cite{Song2019}. This suggests that a proximate Dirac QSL could support a linear coupling between 120$^0$ AFM and FM moments through the monopole effects.

We note also a possible relation to ferromagnetic clusters seen on the Dirac spin liquid candidate samples. These are the compounds in the YCOB-Cl family that show no magnetic order and may be spin liquid candidates, namely those with $x<0.4$ and associated higher density of ABH disorder. Recent work \cite{shivaramPRB2024} found that the unusual magnetization could be explained through ferromagnetic spin clusters arising in the candidate spin liquid regime. These possibly-delocalized clusters can viewed as counterparts of emergent random singlets but with a dominant effect from ferromagnetic interactions that can arise via the interplay of frustrated antiferromagnetic interactions in a strong-disorder RG. It would be interesting to consider how increasing ABH disorder may interplay with a phase transition into a Dirac QSL, and what the role of ferromagnetic spin clusters may be across such a transition.

\section{Acknowledgements} 
Research supported by the 
U.S. Department of Energy (DOE), Office of Science, Basic Energy Sciences (BES), primarily under Early Career Award No. DE-SC0025478 (IK) (theory and computations), and additionally supported by Department of Energy, Office of Science, Basic Energy Sciences Award No. DE-FG02-03ER46076 (PL).
The experimental work at the University of Virginia (BSS and TDF) was supported by U.S.\ National Science Foundation Award DMR-2016909.   SL acknowledges support by the
National Key Research and Development Program of China (Grants No. 2022YFA1403400 and No. 2021YFA1400401), and Chinese Academy of Sciences (Grants No. XDB33000000 and No. GJTD-2020-01). The work at the Naval Research Laboratory (J.P.) has been funded by the Office of Naval Research, under the NRL 6.1 Base Program.\\


\appendix

\begin{figure}[t]
    \centering  \includegraphics[scale=0.4]{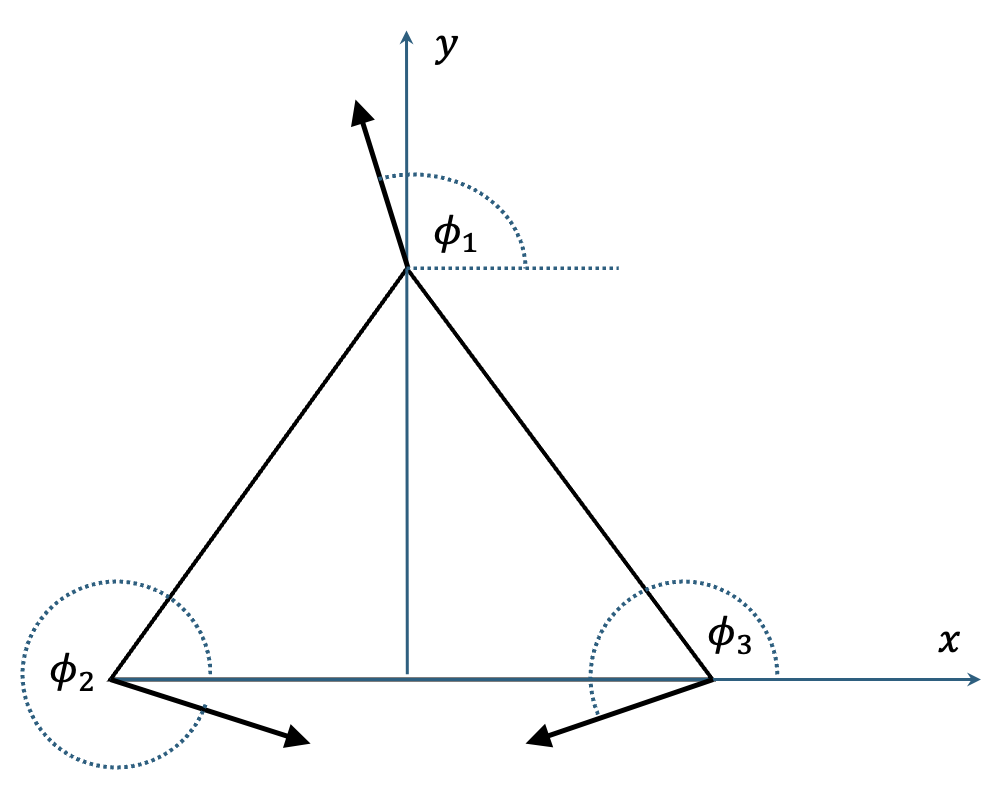}
    \vspace{2.2cm}
    \caption{Reference axes for a single triangle classical analysis. All the angles ($\phi_i$) are measured from horizontal $x-$axis of this image.}
    \label{fig_reference}
\end{figure}

\section{Experimental details}

Here we use single crystals of YCOB-Cl, YCu$_3$(OH)$_6$[(Cl$_x$Br$_{(1-x)}$)$_{3-y}$(OH)$_y$], which are grown using a similar hydrothermal method as described in reference \cite{ZengPRB2022}. This method allows to continuously tune the ratio between Cl and Br. The Cl-doped samples are obtained as hexagonal plates with the in-plane diameter up to 1.3 mm and thickness of 1 mm. 
To remove the possible surface impurities, all the crystals are cleaned ultrasonically in water. Magnetization measurements are performed on a Quantum Design magnetic property measurement system employing a sample holder traversing both halves of the static SQUID detection coils to eliminate any background contribution.

\section{\label{appen_kitaev_ising}In-plane canting due to  Kitaev and bond-Ising terms}

\begin{figure*}
    \centering
    \subfigure[\label{fig_ABH_moment_1}Net moment on triangle for isolated ABH]{
    \includegraphics[scale=0.25]{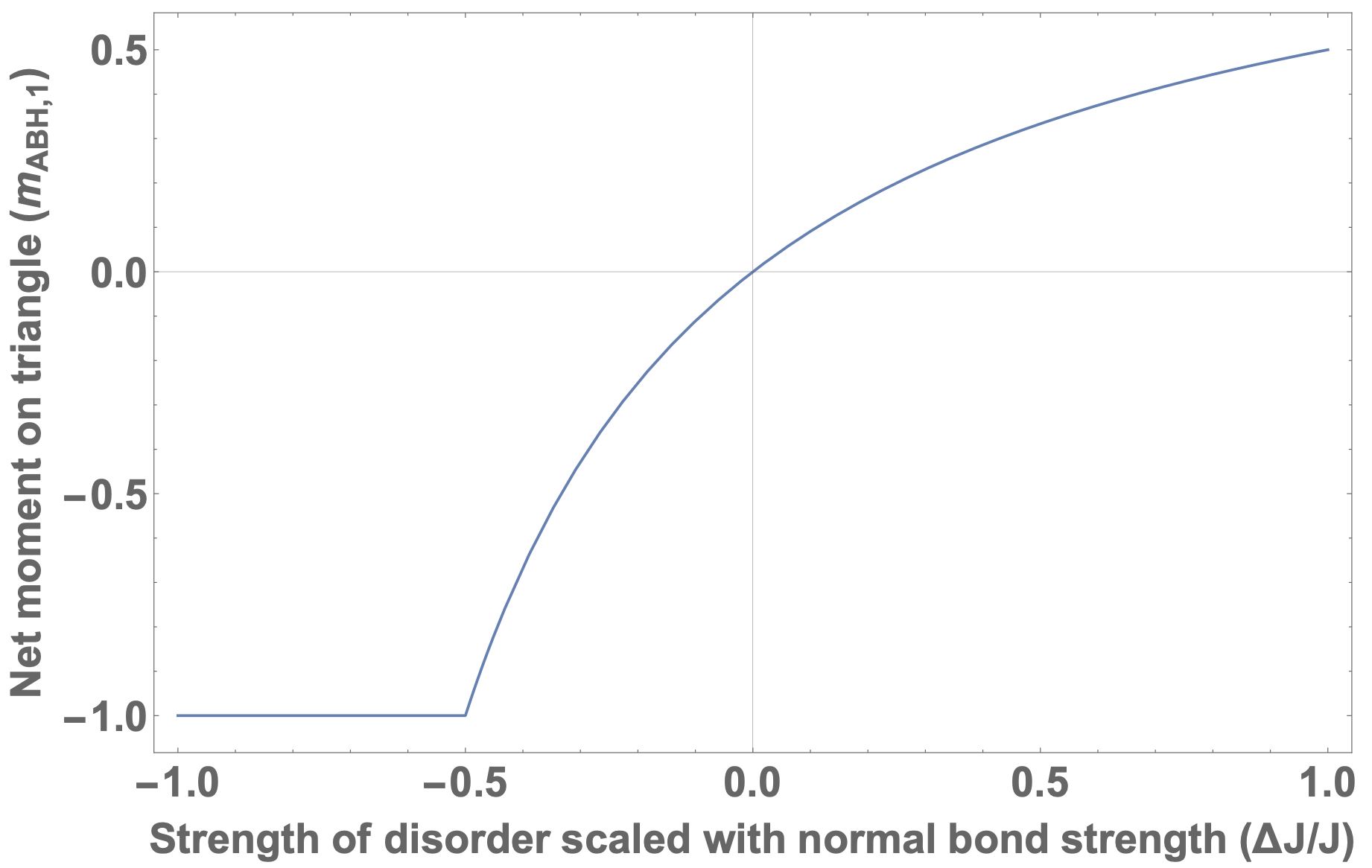}}
    \subfigure[\label{fig_ABH_moment_2}Net moment on triangle for adjacent ABH]{
    \includegraphics[scale=0.25]{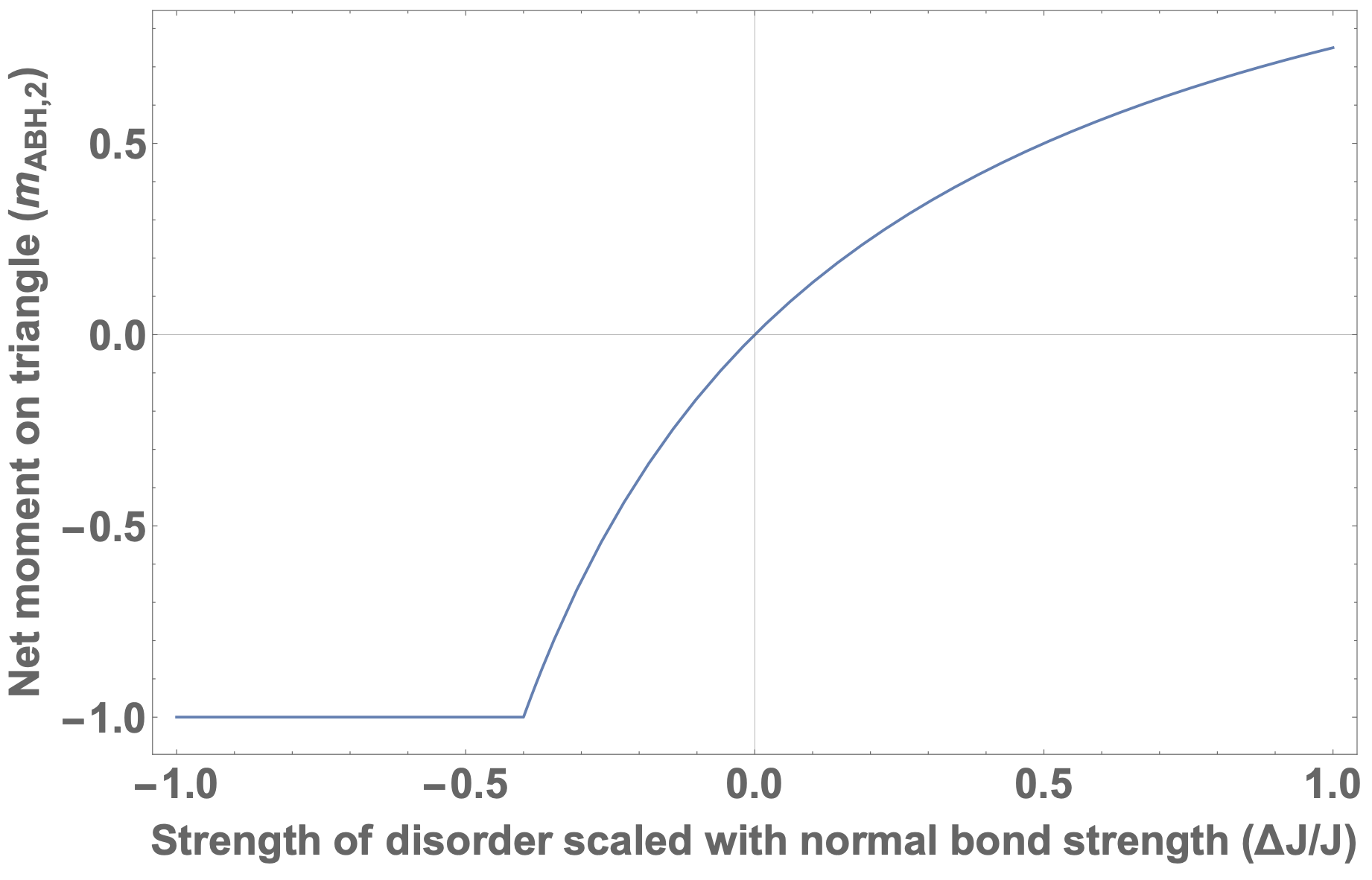}}
    \vspace{2.2cm}
    \caption{Net magnetic moment of a single triangle with ABH disorder: Panel (a) shows net moment on triangles having two bonds of same strength, and one bond different from the others ($H_{{\rm ABH},1}$). (b) shows moment of triangles having all three bond strength different from each other ($H_{{\rm ABH},2}$). Panel (b) occurs if two ABH are adjacent. We find that panel (b) is not qualitatively different from panel (a), so in the case where most ABH are dilute, getting some adjacent ABH does not give any qualitatively new effects, and we can safely ignore it below.}
    \label{fig_net moment_ABH_single triangle classical}
\end{figure*}

\begin{figure*}
    \centering
    \subfigure[\label{fig_canting_ABH_sttp} Various spin orientations and associated canting for strong bond on top ABH]{
    \includegraphics[scale=0.5]{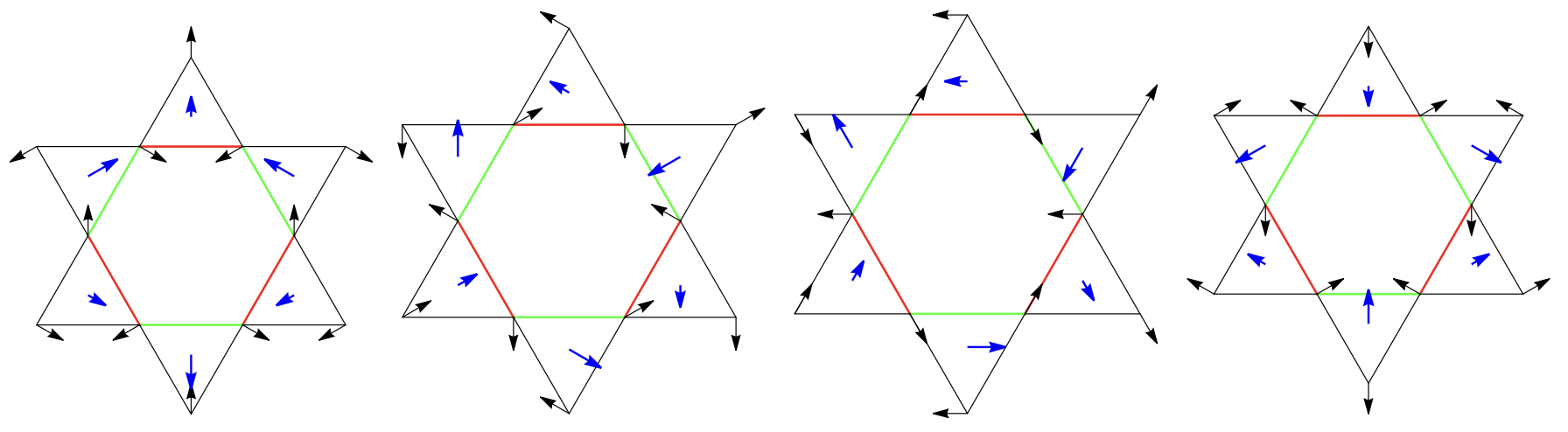}
    }
    \subfigure[\label{fig_canting_ABH_wktp}Various spin orientations and associated canting for weak bond on top ABH]{
    \includegraphics[scale=0.5]{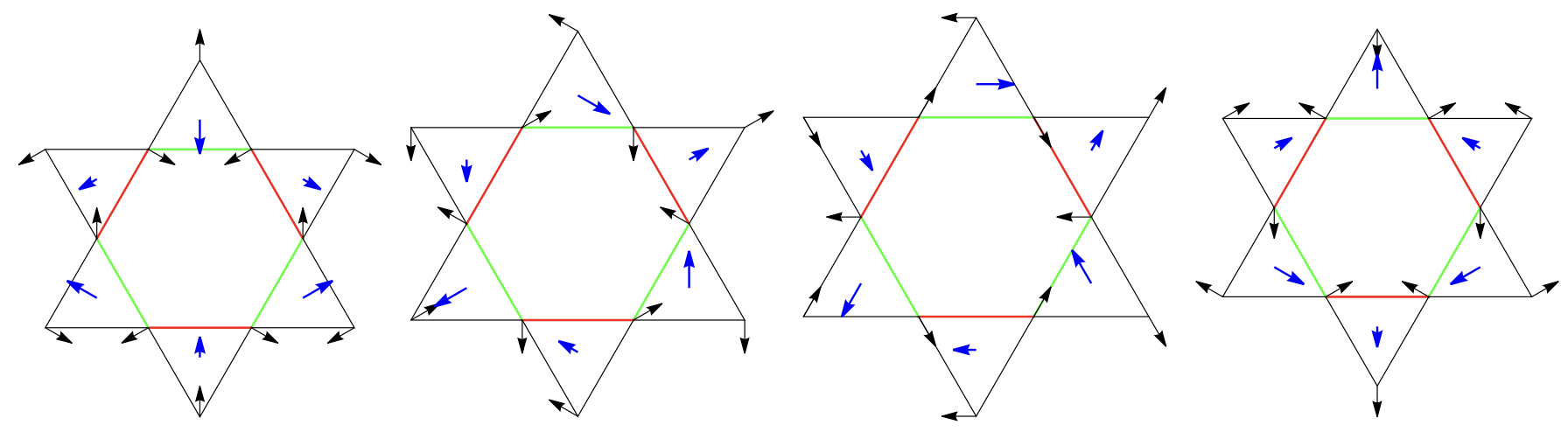}
    }
    \vspace{2.2cm}
    \caption{Canting due to the ABH disorder: Red and green bonds are stronger ($J+\Delta J$) and weaker ($J-\Delta J$) bonds, respectively with the choice of $\Delta J/J=0.25$ for all figures. The black arrows on the kagome vertices represent the spins of the corresponding antichiral order. The blue arrows on the centers of each triangles represent the net moment of the corresponding triangles. Panels (a) and (b) show two different possibilities of ABH with red or green bond being on the top of the hexagon, respectively. The two configurations are related by a $C_3$ symmetry. Four figures on each row show different configurations due to $\phi_1=~\pi/2,~5\pi/6,~\pi,~3\pi/2$. Note that the blue arrows (net moments) always forms a vortex configuration with vorticity -1.}
    \label{fig_canting_ABH}
\end{figure*}

Here we give further details of two potential perturbations to the Heisenberg and out-of-plane DM Hamiltonian which are Kitaev interaction and bond-Ising  interaction. 
Among them, as we argued in the main text (see Sec.\ \ref{sec_antichiral}), microscopically the dominant contribution to canting comes from the Kitaev term. 
Both of these interactions have similar symmetry, and they break continuous $U(1)$ symmetry of the Heisenberg and DM term. 
Via a single triangle (${ q}=0$) classical analysis, we show that such terms indeed favor a uniform FM canting of the underlying antichiral AFM order.

Since we are considering q=0 magnetic orders, it is enough to look at a single triangle. 
The bond-Ising and Kitaev terms are given by:
\begin{widetext}
\begin{align}
   & H_\text{Ising}=J_\text{Ising}\sum_{\langle ij\rangle} \left(S_i\cdot ({\bf r_i}-{\bf r_j})\right)\left(S_j\cdot ({\bf r_i}-{\bf r_j})\right),\\
   & H_\text{Kitaev}=J_\text{Kitaev}\sum_{\langle ij\rangle_\mu}S_i^\mu S_j^\mu~~(\mu=X,Y,Z),\\
    &X = \Big\{-\frac{1}{\sqrt{2}}, -\frac{1}{\sqrt{6}}, \frac{1}{\sqrt{3}}\Big\},
    ~Y = \Big\{\frac{1}{\sqrt{2}}, -\frac{1}{\sqrt{6}}, \frac{1}{\sqrt{3}}\Big\},
    ~Z = \Big\{0,\sqrt{\frac{2}{3}}, \frac{1}{\sqrt{3}}\Big\},
\end{align}
where  in the last line $X, Y$ and $Z$ of the Kitaev interaction are given in terms of the reference axes $(x, y, z)$ described in Fig. \ref{fig_reference}. We note that although both the terms break the continuous $U(1)$ symmetry,  there is still an emergent $U(1)$ symmetry in the antichiral subspace at least at the classical level. Here we perturbatively compute the classical energy of the above two Hamiltonian around the anti-chiral state. For this we define,
\begin{align}
    &\phi_2-\phi_3=\frac{2\pi}{3}+\frac{a}{\sqrt{3}}~~, ~~\phi_2+\phi_3-2\phi_1=2\pi+b.
    \label{eq_expansion around antichiral_ising2}
\end{align}
Note that the scaling of $a$ and $b$ is chosen such that when Eq. \ref{eq_expansion around antichiral_ising2} is substituted in the classical Heisenberg energy, the leading order correction becomes $\propto (a^2+b^2)$. Solving $\phi_2$ and $\phi_3$ from the above two equations, we can substitute them in the classical energy expression for the bond-Ising and Kitaev  terms. Expanding in powers of $a$ and $b$, we obtain (here we express the classical energies in the units of the corresponding echange interaction scale):

\begin{align}
    \frac{H_\text{Ising}}{J_{\text{Ising}}}&=\frac{1}{4}\left[2\cos\left(2\phi_1+b\right)-\sqrt{3}\cos\left(2\phi_1+\frac{b}{2}+\frac{a}{2\sqrt{3}}-\frac{\pi}{6}\right)\right.\nonumber\\
    &-\cos\left(2\phi_1+\frac{b}{2}-\frac{a}{2\sqrt{3}}-\frac{2\pi}{3}\right)+2\sin\left(\frac{a}{2\sqrt{3}}-\frac{b}{2}-\frac{\pi}{6}\right)+2\sin\left(\frac{a}{2\sqrt{3}}+\frac{b}{2}-\frac{\pi}{6}\right)\nonumber\\
    &\left.-2\sin\left(\frac{a}{\sqrt{3}}+\frac{\pi}{6}\right)-\sin\left(2\phi_1+\frac{a}{2\sqrt{3}}+\frac{b}{2}-\frac{\pi}{6}\right)-\sqrt{3}\sin\left(2\phi_1-\frac{a}{2\sqrt{3}}+\frac{b}{2}+\frac{2\pi}{3}\right)\right]\\
    &\approx-\frac{3}{4}-\frac{1}{4} \left(a\cos(2\phi_1)+b\sin(2\phi_1)\right)+\mathcal{O}(a^2,b^2,ab),\\
    \frac{H_\text{Kitaev}}{J_\text{Kitaev}}&=\frac{1}{6} \left(-2 \cos\left(b + 2 \phi_1\right) + 
   \sqrt{3} \cos\left(\frac{1}{6} \left(\sqrt{3} a + 3 b - \pi + 12 \phi_1\right)\right) \right.\nonumber\\
      &+ 
   \cos\left(\frac{1}{6} \left(-\sqrt{3} a + 3 b + 4 \pi + 12 \phi_1\right)\right) + 
   2 \sin\left(\frac{1}{6} \left(\sqrt{3} a - 3 b - \pi\right)\right) + 
   2 \sin\left(\frac{1}{6} \left(\sqrt{3} a + 3 b - \pi\right)\right)\nonumber\\
   &\left. - 
   2 \sin\left(\frac{1}{\sqrt{3}} a + \frac{\pi}{6}\right) + 
   \sin\left(\frac{1}{6} \left(\sqrt{3} a + 3 b - \pi + 12 \phi_1\right)\right) + 
   \sqrt{3} \sin\left(\frac{1}{6} \left(-\sqrt{3} a + 3 b + 4 \pi + 12 \phi_1\right)\right)\right)\\
   &\approx -\frac{1}{2}+\frac{1}{6}\left(a\cos(2\phi_1)+b\sin(2\phi_1)\right)+\mathcal{O}(a^2,b^2,ab).
\end{align}

Note that $\phi_1$ independent constant term is the consequence of the emergent $U(1)$ invariance in the antichiral subspace.
Substituting: 
\begin{align}
    a=r\cos(f_1)~~, ~~b=r\sin(f_1)
\end{align}
we get:
\begin{align}
    &H_\text{Ising}/J_\text{Ising}\approx-\frac{3}{4}-\frac{1}{4}r\cos(f_1-2\phi_1)+\mathcal{O}(a^2,b^2,ab),\\
    &H_\text{Kitaev}/J_\text{Kitaev}\approx-\frac{1}{2}+\frac{1}{6}r\cos(f_1-2\phi_1)+\mathcal{O}(a^2,b^2,ab).
\end{align}
Note that the classical energy is minimized by,
\begin{align}
    &J_\text{Ising}<0 ~(FM)~: ~f_1= 2\phi_1+\pi~~,~~ J_\text{Ising}>0 ~(AFM)~: ~f_1= 2\phi_1\\
    &J_\text{Kitaev}>0 ~(AFM)~: ~f_1= 2\phi_1+\pi~~,~~ J_\text{Kitaev}<0 ~(FM)~: ~f_1= 2\phi_1
\end{align}
The Ising and Kitaev term give rise to opposite kind of canting. Note that the $U(1)$ symmetry breaking decides the canting angle of the spin, hence the net canting moment on the triangle. The net canted moment on the triangle is given by,

\begin{align}
    m=\Bigg\{\begin{array}{c}
    \frac{1}{2}\left\{-\cos(\phi_1)~,~\sin(\phi_1)\right\}~,~~\text{for FM Ising and AFM Kitaev.}\\
    \frac{1}{2}\left\{\cos(\phi_1)~,~-\sin(\phi_1))\right\}~,~~\text{for AFM Ising and FM Kitaev.}
    \end{array}
\end{align}
    
\end{widetext}

We can parametrize the canted moment as,
\begin{align}
    m = \frac{1}{2}\left\{\cos(\gamma)~,~\sin(\gamma)\right\}.
\end{align}
Then we find that for AFM (FM) bond-Ising (Kitaev),  $\gamma = - \phi_1$ and FM (AFM) bond-Ising (Kitaev) gives time-reversed canted moments, $\gamma = - \phi_1 + \pi$. Note that $\phi_1$ is the global choice of the antichiral order which breaks the $U(1)$ symmetry of the Heisenberg model and DM interaction. In Fig. \ref{fig_canting} of the main text, we plot the net FM moments as computed above on a star of David of the kagome lattice for different choices of $\phi_1$. Note that although both terms generate similar canting, the magnitude of their effect depends on the coupling constant of these two terms. Generically, for the kagome lattice with DM anisotropy, it is expected that Kitaev interaction is more dominant than the other anisotropic exchanges such as in-plane bond-Ising. Therefore, in the main text, we only discuss the Kitaev interaction leaving aside the contribution from bond-Ising term.

\section{\label{appen_chiral}Lack of in-plane canting for chiral state}
Here we give additional details for why the chiral (as opposed to antichiral) state does not get in-plane canting. 

First recall the discussion from the main text. The DM interaction appears in the first order of spin-orbit coupling ($\lambda_{\rm SOC}$), whereas the Kitaev and bond-Ising terms appear in the second order in  $\lambda_{\rm SOC}$. So DM interactions whether in-plane or out-of-plane are larger than the anisotropic Ising interactions. Therefore, if there was a chiral order in the material, the symmetry allowed in-plane already gives out-of-plane ferromagnetism, which is not observed in the experiments. This clearly rules out the possibility of chiral order, as opposed to antichiral order. The in plane FM order requires antichiral order, and an additional ingredient beyond DM.

Now keeping the above in mind we can separately ask what does the in plane anisotropy such as Ising terms do to the chiral order. The main difference in the chiral case compared to the antichiral is there is no emergent $U(1)$ symmetry in the chiral subspace for the bond-Ising or Kitaev interaction, therefore a chiral state is chosen by the anisotropic interactions. If we do the perturbation analysis of energy gained by canting about this state, $\phi_2 \rightarrow \frac{2 \pi}{3} + \phi_1 + \frac{\left(b + a/\sqrt{3}\right)}{2}, \phi_3 \rightarrow \frac{4 \pi}{3} + \phi_1 + \frac{\left( b - a/\sqrt{3}\right)}{2}$ , we obtain:
\begin{align}
    &\left(\frac{H_\text{Ising}}{J_{\text{Ising}}}\right)_\text{chiral}
    \approx-\frac{3}{4} \left(1 + 2 \cos\left(2 \phi_1\right)\right) + b \sin\left(2 \phi_1\right)+...\\
    &\left(\frac{H_\text{Kitaev}}{J_\text{Kitaev}}\right)_\text{chiral} \approx -\frac{1}{2} - \cos\left(2 \phi_1\right) + \frac{2}{3} b \sin\left(2 \phi_1\right)+...
\end{align}
The minima is obtained by setting $\phi_1=0$.
Clearly, the the energy gain due to canting is zero to the  first order in perturbation. Hence the chiral order does not get any  in-plane canting to the leading order.

\section{\label{appen_ABH}Classical analysis of ABH: details of classical mechanism relieving frustration, and canting due to strong bond disorder}

In this section, we describe the details of classical mechanism relieving frustration, and local canting due to ABH disorder, from the ${ q}=0$ single triangle classical analysis.

Note that, depending on the density of the ABH, there can be two different disordered triangles possible: triangles with one of the bond  either stronger or weaker than the other two (occurs even for a single isolated ABH), and triangles with all the three bonds are of different strength (occurs only for adjacent ABH). The spin Hamiltonians for these two cases are given by:
\begin{align}
    &H_{{\rm ABH},1}=H_{\text{Heisenberg}}+H_\text{DMz}+\Delta J S_2\cdot S_3,\\
    &H_{{\rm ABH},2}=H_{\text{Heisenberg}}+H_\text{DMz}+\Delta J\left( S_2\cdot S_3- S_1\cdot S_2\right).
\end{align}
We mostly focus on the first case (dilute ABH limit) but for completeness also mention the second case.

For reference of the labels of the spins, see Fig. \ref{fig_reference}. We note that both of the above Hamiltonian do not break global $U(1)$ invariance. Hence we measure the angle of $S_2$ and $S_3$ from $S_1$, which we denote as $\tilde{\phi}_2$ and $\tilde{\phi}_3$. The antichiral AFM order (atleast short ranged) is established by the combination of $H_\text{Heisenberg}$ and $H_\text{DMz}$, before the onset of the FM canting. Therefore,  we treat the disorder term perturbatively  around the antichiral states.  We assume:
\begin{align}
    \tilde{\phi}_2=\frac{4\pi}{3}+\delta_{ABH},~~ \tilde{\phi}_3=\frac{2\pi}{3}-\delta_{ABH},
\end{align}
where $\delta_{ABH}$ denotes the amount of FM canting due to ABH disorder(see Fig. \ref{fig_reference}). Substituting the above in  the classical energy expression, and solving their minima,  we obtain:
\begin{widetext}
\begin{itemize}
    \item  For $H_{ABH,1}$ (isolated ABH):
\begin{align}
    \delta_{ABH,1}=\Bigg\{\begin{array}{lr}
         -\frac{\pi}{3}  &~~\text{for} ~~\Delta J<-0.5\\
          i\left(\log_e(4)-\log_e\left(\frac{(i+\sqrt{3})\left(-i-i\Delta J+\sqrt{(1+\Delta J)^2(3+4\Delta J(2+\Delta J))}\right)}{2(1+\Delta J)^2}\right)\right)  &~~\text{for} ~~\Delta J>-0.5
    \end{array}
\end{align}
The net moment of the triangle can be computed as:
\begin{align}
    m_{ABH,1}=1+2\cos\left(\frac{4\pi}{3}+\delta_{ABH,1}\right)=\Bigg\{\begin{array}{lr}
         -1  &~~\text{for} ~~\Delta J<-0.5\\
         \frac{\Delta J}{1+\Delta J} &~~\text{for} ~~\Delta J>-0.5
    \end{array}
\end{align}
Calculating the energy for such canted state we obtain:
\begin{align}
    E_{ABH,1}\approx-\frac{3J}{2} - \frac{\Delta J}{2} - \frac{(\Delta J)^2}{2J}
\end{align}
The first order term cancels out when averaged over  triangles with strong and weak bond, leaving only the second order term. So the energy per bond is modified 
from $J/2$ to $J/2 - (\Delta J)^2/6J$.

    \item  For $H_{ABH,2}$ (adjacent ABH):
\begin{align}
    \delta_{ABH,2}=\Bigg\{\begin{array}{lr}
         -\frac{\pi}{3}  &~~\text{for} ~~\Delta J<-0.4\\
          \tan^{-1}\left(\frac{\sqrt{3}\left(-2+\Delta J(\Delta J-1)+\sqrt{(1+\Delta J)^2(2+\Delta J)(2+5\Delta J)}\right)}{\left(2+\Delta J-(\Delta J)^2+3\sqrt{(\Delta J+1)^2(\Delta J+2)(5\Delta J+2)}\right)}\right)  &~~\text{for} ~~\Delta J>-0.4
    \end{array}
\end{align}
The net moment of the triangle is given by:
\begin{align}
    m_{ABH,2}=1+2\cos\left(\frac{4\pi}{3}+\delta_{ABH,2}\right)=\Bigg\{\begin{array}{lr}
         -1  &~~\text{for} ~~\Delta J<-0.4\\
         \frac{3\Delta J}{2+2\Delta J} &~~\text{for} ~~\Delta J>-0.4
    \end{array}
\end{align}
\end{itemize}
    
\end{widetext}

Note that the qualitative results obtained in the above two cases are very similar. The difference lies in the detailed values of the canted moment. In Fig. \ref{fig_net moment_ABH_single triangle classical}, we plot the net FM moment of the triangle for the both the kinds of disorder. 
In Fig. \ref{fig_canting_ABH}, we show the FM canting of the triangles surrounding the ABH (star of David). Note that due to a $C_3$ symmetry of the ABH hexagon, the canted configuration also has a $C_3$ rotation symmetry, leading to zero net moment on each such hexagons. Interestingly, the canted moments due to ABH forms a vortex configuration with voricity -1, which is a consequence of the underlying antichiral pattern.


\begin{thebibliography}{0}%
\makeatletter
\providecommand \@ifxundefined [1]{%
 \@ifx{#1\undefined}
}%
\providecommand \@ifnum [1]{%
 \ifnum #1\expandafter \@firstoftwo
 \else \expandafter \@secondoftwo
 \fi
}%
\providecommand \@ifx [1]{%
 \ifx #1\expandafter \@firstoftwo
 \else \expandafter \@secondoftwo
 \fi
}%
\providecommand \natexlab [1]{#1}%
\providecommand \enquote  [1]{``#1''}%
\providecommand \bibnamefont  [1]{#1}%
\providecommand \bibfnamefont [1]{#1}%
\providecommand \citenamefont [1]{#1}%
\providecommand \href@noop [0]{\@secondoftwo}%
\providecommand \href [0]{\begingroup \@sanitize@url \@href}%
\providecommand \@href[1]{\@@startlink{#1}\@@href}%
\providecommand \@@href[1]{\endgroup#1\@@endlink}%
\providecommand \@sanitize@url [0]{\catcode `\\12\catcode `\$12\catcode `\&12\catcode `\#12\catcode `\^12\catcode `\_12\catcode `\%12\relax}%
\providecommand \@@startlink[1]{}%
\providecommand \@@endlink[0]{}%
\providecommand \url  [0]{\begingroup\@sanitize@url \@url }%
\providecommand \@url [1]{\endgroup\@href {#1}{\urlprefix }}%
\providecommand \urlprefix  [0]{URL }%
\providecommand \Eprint [0]{\href }%
\providecommand \doibase [0]{https://doi.org/}%
\providecommand \selectlanguage [0]{\@gobble}%
\providecommand \bibinfo  [0]{\@secondoftwo}%
\providecommand \bibfield  [0]{\@secondoftwo}%
\providecommand \translation [1]{[#1]}%
\providecommand \BibitemOpen [0]{}%
\providecommand \bibitemStop [0]{}%
\providecommand \bibitemNoStop [0]{.\EOS\space}%
\providecommand \EOS [0]{\spacefactor3000\relax}%
\providecommand \BibitemShut  [1]{\csname bibitem#1\endcsname}%
\let\auto@bib@innerbib\@empty
\end{thebibliography}%


\begin{thebibliography}{99}

\bibitem{NormanRMP2016} M. R. Norman, Colloquium:  Herbertsmithite and the search for the quantum spin liquid, Rev. Mod. Phys. {88}, 041002 (2016). 



\bibitem{sunJCHEM2016} W. Sun,   Y.-X. Huang, S. Nokhrin,  Y. Pan,  and  J.-X. Mi, Perfect Kagomé lattices in YCu$_3$(OH)$_6$Cl$_3$: a new candidate for the quantum spin liquid state, J. Mater. Chem. C  {4}, 8772-8777 (2016).


\bibitem{ZorkoPRB2019}A. Zorko, M. Pregelj, M. Klanjšek, M. Gomilšek, Z. Jagličić, J. S. Lord, J. A. T. Verezhak, T. Shang, W. Sun, and J.-X. Mi, Coexistence of magnetic order and persistent spin dynamics in a quantum kagome antiferromagnet with no intersite mixing, Phys. Rev. B  {99}, 214441 (2019). 


\bibitem{ZorkoPRB2019negative} A. Zorko , M. Pregelj, M. Gomilšek, M. Klanjšek, O. Zaharko, W. Sun, and J.-X. Mi, Negative-vector-chirality 120$^0$ spin structure in the defect- and distortion-free quantum kagome antiferromagnet YCu$_3$(OH)$_6$Cl$_3$, Phys. Rev. B  {100}, 144420 (2019). 


\bibitem{ArhPRL2020}T. Arh, M. Gomilšek, P. Prelovšek, M. Pregelj, M. Klanjšek, A. Ozarowski, S. J. Clark, T. Lancaster, W. Sun, J.-X. Mi, and A. Zorko, Origin of Magnetic Ordering in a Structurally Perfect Quantum kagome Antiferromagnet, Phys. Rev. Lett.  {125}, 027203 (2020). 

\bibitem{BernuPRB2020}B. Bernu, L. Pierre, K. Essafi, and L. Messio, Effect of perturbations on the kagome $S=\frac{1}{2}$ antiferromagnet at all temperatures, Phys. Rev. B  {101}, 140403 (2020). 

\bibitem{Messio2010} L. Messio, O. Cépas, and C. Lhuillier, Schwinger-boson approach to the kagome antiferromagnet with Dzyaloshinskii Moriya interactions:Phase diagram and dynamical structure  factors, Phys. Rev. B  {81}, 064428 (2010).  

\bibitem{Cepas2008}O. Cépas, C. M. Fong, P. W. Leung, and C. Lhuillier, Quantum  phase transition induced by Dzyaloshinskii-Moriya interactions in the kagome antiferromagnet, Phys. Rev. B  {78}, 140405(R) (2008).  

\bibitem{HeringPRB2017}M. Hering and J. Reuther, Functional renormalization group analysis of Dzyaloshinsky-Moriya and Heisenberg spin interactions on the kagome lattice, Phys. Rev. B  {95}, 054418 (2017). 

\bibitem{BarthelemyPRM2019} Q. Barthélemy, P. Puphal, K. M. Zoch, C. Krellner, H. Luetkens, C. Baines, D. Sheptyakov, E. Kermarrec, P. Mendels, and F. Bert, Local study of the insulating quantum kagome antiferromagnets, Phys Rev. Materials  {3}, 074401 (2019). 


\bibitem{HongPRB2022} X. Hong, M. Behnami, L. Yuan, B. Li, W. Brenig, B. Büchner, Y. Li, and C. Hess, Heat transport of the kagome Heisenberg quantum spin liquid candidate YCu$_3$(OH)$_{6.5}$Br$_{2.5}$: Localized magnetic excitations and a putative spin gap, Phys. Rev. B  {106}, L220406 (2022). 

\bibitem{Lunature2022}F. Lu, L. Yuan, J. Zhang, B. Li, Y. Luo, and  Y. Li, The observation of quantum fluctuations in a kagome Heisenberg antiferromagnet, Commun. Phys.  {5}, 272  (2022). 

\bibitem{ChenJMMM2020}X.-H. Chen, Y.-X. Huang, Y. Pan, and J.-X. Mi, Quantum spin liquid candidate With an almost perfect kagome layer,  YCu$_3$(OH)$_6$Br$_2$[Br$_x$(OH)$_{(1-x)}$],  Journal of
Magnetism and Magnetic Materials   {512}, 167066 (2020). 



\bibitem{ZengPRB2022} Z. Zeng, X. Ma, S. Wu, H.-F. Li, Z. Tao, X. Lu, X.-H. Chen, J.-X. Mi, S.-J. Song, G.-H. Cao, G. Che, K. Li, G. Li, H. Luo, Z. Y. Meng, and S. Li, Possible Dirac quantum spin liquid in the kagome quantum antiferromagnet YCu$_3$(OH)$_6$Br$_2$[Br$_x$(OH)$_{(1-x)}$], Phys. Rev. B  {105}, L121109 (2022). 


\bibitem{suetsugu_gapless_2024} S. Suetsugu, T. Asaba, S. Ikemori, Y. Sekino, Y. Kasahara, K. Totsuka, B. Li, Y. Zhao, Y. Li, Y. Kohama, and Y. Matsuda, Gapless spin excitations in a quantum spin liquid state of S=1/2 perfect kagome antiferromagnet, arXiv:2407.16208 (2024).




\bibitem{Xu2023} A. Xu,  Q. Shen, B. Liu, Z. Zeng, L. Han, L. Yan, J. Luo, J. Yang, R. Zhou, and S. Li, Magnetic ground states in the kagome system YCu$_3$(OH)$_6$ [(Cl$_x$Br$_{1-x}$)$_{3-y}$(OH)$_y$],  Phys. Rev. B  {110}, 085146 (2024).  


\bibitem{shivaramPRB2024} B. S. Shivaram, J. C. Prestigiacomo, A. Xu, Z. Zeng, T. D. Ford, I. Kimchi, S. Li, and P. A. Lee, Nonanalytic magnetic response and intrinsic ferromagnetic clusters in a kagome spin-liquid candidate, Phys. Rev. B  {110}, L121105 (2024).


\bibitem{Liu2022} J. Liu, L. Yuan, X. Li, B. Li, K. Zhao, H. Liao, and Y. Li, Gapless spin liquid behavior in a kagome Heisenberg antiferromagnet with randomly distributed hexagons of alternate bonds, Phys. Rev. B  {105}, 024418 (2022). 

\bibitem{Zengpreprintneutron} Z. Zeng, C. Zhou, H. Zhou, L. Han, R. Chi, K. Li, M. Kofu, K. Nakajima, Y. Wei, W. Zhang, D. G. Mazzone, Z. Y. Meng, and S. Li,
Spectral evidence for Dirac spinons in a kagome lattice antiferromagnet,
Nature Physics 20, 1097–1102 (2024).


\bibitem{RanPRL2007} Y. Ran, M. Hermele, P. A. Lee, and X.-G. Wen, Projected-Wave-Function Study of the Spin-1/2 Heisenberg Model on the Kagomé Lattice, Phys. Rev. Lett.  {98}, 117205 (2007).


\bibitem{HermelePRB2008} M. Hermele, Y. Ran, P. A. Lee, and X.-G. Wen, Properties of an algebraic spin liquid on the kagome lattice, Phys. Rev. B  {77}, 224413 (2008).



\bibitem{Puphal2017}P. Puphal, M. Bolte, D. Sheptyakov, A. Pustogow, K. Kliemt, M. Dressel, M. Baenitz, and C. Krellner, Strong magnetic frustration in Y$_3$Cu$_9$(OH)$_{19}$Cl$_8$: A distorted kagome antiferromagnet, J. Mater. Chem. C  {5}, 2629 (2017).   


\bibitem{ChatterjeePRB2023} D. Chatterjee, P. Puphal, Q. Barthélemy, J. Willwater, S. Süllow, C. Baines, S. Petit, E. Ressouche, J. Ollivier, K. M. Zoch, C. Krellner, M. Parzer, A. Riss, F. Garmroudi, A. Pustogow, P. Mendels, E. Kermarrec, and F. Bert, From spin liquid to magnetic ordering in the anisotropic kagome Y-kapellasite Y$_3$Cu$_9$(OH)$_{19}$Cl$_8$: A single-crystal study, Phys. Rev. B  {107}, 125156 (2023). 




\bibitem{kimchi_kitaev_2014} I. Kimchi  and A. Vishwanath,
Kitaev-Heisenberg models for iridates on the triangular, hyperkagome, kagome, fcc, and pyrochlore lattices, Phys. Rev. B  {89}, 014414 (2014). 


\bibitem{ElhajalPRB2002}M. Elhajal, B. Canals, and C. Lacroix, Symmetry breaking due to Dzyaloshinsky-Moriya interactions in the kagome lattice, Phys. Rev. B  {66}, 014422 (2002).  

\bibitem{lee2018} C.-Y. Lee, B. Normand, and Y.-J. Kao, Gapless spin liquid in the kagome Heisenberg antiferromagnet with Dzyaloshinskii-Moriya interactions, 
Phys. Rev. B  {98}, 224414 (2018).


\bibitem{moriya_anisotropic_1960} T. Moriya,  Anisotropic Superexchange Interaction and Weak Ferromagnetism, Phys. Rev.  {120}, 91 (1960). 


\bibitem{Vojta2013} T. Vojta, Phases and phase transitions in disordered quantum systems, AIP Conf. Proc.  {1550}, 188–247 (2013).


\bibitem{Song2019} X.-Y. Song, C. Wang, A. Vishwanath, and Y.-C. He, Unifying description of competing orders in two-dimensional quantum magnets, Nat. Commun.  {10}, 4254 (2019).  








\end {thebibliography}
\end{document}